\documentclass[review, 12pt]{elsarticle}

\usepackage{amssymb}
\usepackage{bm}
\usepackage[colorlinks=true, linkcolor=red, anchorcolor=blue,citecolor=green]{hyperref}
\usepackage{amsmath}
\usepackage{caption}
\usepackage{subcaption}

\usepackage{cleveref}
\usepackage{multirow}
\usepackage{color,soul}
\usepackage[table]{xcolor}

\usepackage[switch]{}
\usepackage{lineno}

\newcommand\K{{\bf k}}

\newcommand\U{{\bf u}}

\newcommand\X{{\bf x}}

\newcommand\be{\begin{equation}}
\newcommand\nd{\end{equation}}
\newcommand\bed{\begin{displaymath}}
\newcommand\ndd{\end{displaymath}}

\newcommand\ba{\begin{array}}
\newcommand\ea{\end{array}}
\newcommand\bea{\begin{eqnarray}}
\newcommand\nda{\end{eqnarray}}

\begin{document}
% \captionsetup[figure]{labelfont={bf},name={Fig.},labelsep=period}
% \captionsetup[table]{labelfont={bf},name={Table.},labelsep=period}
\sethlcolor{yellow}
\setstcolor{red}
\soulregister\cite7 % for \cite
\soulregister\ref7
%\UseRawInputEncoding

\begin{frontmatter}

\title{An Unsplit Scheme for Interface Advection in the Edge-based Interface Tracking (EBIT) Method}

\author[a]{Jieyun Pan\corref{cor1}}
\ead{yi.pan@sorbonne-universite.fr}

\author[a,b]{Tian Long}
\ead{tian.long@dalembert.upmc.fr}

\author[c]{Ruben Scardovelli}
\ead{ruben.scardovelli@unibo.it}

\author[a,d]{St\'{e}phane Zaleski}
\ead{stephane.zaleski@sorbonne-universite.fr}

\cortext[cor1]{Corresponding author}
\address[a]{Sorbonne Universit\'{e} and CNRS, Institut Jean Le Rond d'Alembert UMR 7190, F-75005 Paris, France}
\address[b]{School of Aerospace, Xi'an Jiaotong University, Xi'an, 710049, PR China}
\address[c]{DIN - Lab. di Montecuccolino, Universit\`a di Bologna, I-40136 Bologna, Italy}
\address[d]{Institut Universitaire de France, Paris, France}
\begin{abstract}

We propose an unsplit scheme for interface advection in a novel
Front-Tracking method, called the Edge-Based Interface Tracking (EBIT) method. 
In the EBIT method, the markers are placed on the grid edges, and their
connectivity is implicitly represented using a color vertex field.
These local features enable almost automatic parallelization and
simplify marker addition or removal compared to traditional Front-Tracking methods.

In our previous publications, a split scheme was used to advect the
interface in the EBIT method. Although the split scheme facilitates the
extension of the EBIT method to three dimensions, it also imposes some
limitations. First, implementing high-order time integration methods for
interface advection becomes challenging.
Second, in multiscale simulations using Front-Tracking methods with a
boundary layer model applied to the interface, the redistribution of physical
quantities inside the boundary layer, such as a scalar concentration, 
along the interface becomes difficult due to the loss of correspondence
between interface segments before and after each advection step, caused 
by the operator split.

Here, we present an unsplit scheme for interface advection to enhance the 
capability of the EBIT method. In the unsplit scheme, a reconstruction step
based on a circle fit, similar to that used in the split scheme, is performed
to reposition markers on the cell edges. Furthermore, the algorithm used 
to update the color vertex field, which implicitly represents marker connectivity and distinguishes among ambiguous topology configurations, is modified to achieve consistent connectivity results.

An EBIT method based on an unsplit scheme has been implemented in the free
Basilisk platform, and validated with four kinematic test cases: translation
with uniform velocity,  solid body rotation,  Zalesak's disk rotation, and
single vortex test. The results are compared with those obtained using the
original EBIT method based on a split scheme.
\end{abstract}


\end{frontmatter}
%\linenumbers

% \linenumbers

% main text
\section{Introduction}
Multiphase flows involve a wide range of spatial scales and are present in many 
natural phenomena and engineering applications. Some examples include breaking waves,
atomizing jets, as well as boiling and condensation in heat exchangers.
The direct numerical simulation of such flows remains a formidable challenge due to
the complexity of physical modeling and numerical schemes.

An accurate prediction of the interface motion is one of the most crucial issues
in the field of multiphase flow simulation. It can be broadly divided into two
problems: the kinematic problem, which determines the motion of the interface 
separating the fluid phases, given the velocity field and the rate of phase change,
and the dynamic problem, which solves the momentum and energy conservation equations,
given the fluid properties.
For the kinematic problem, several methods have been developed that
are generally classified as Front-Capturing methods, including
Volume-of-Fluid (VOF)
\cite{Hirt_1981_39, Brackbill_1992_100, Lafaurie_1994_113, Scardovelli_1999_31},
Level-Set (LS) \cite{Osher_1988_79, Osher_2001_169},
and Front-Tracking methods \cite{Unverdi_1992_100, Tryggvason_2011_book}. 

In Front-Capturing methods, a tracer or marker function $f   (\X,t)$ 
is integrated in time with a prescribed velocity field $\U(\X,t)$.
The function $f$ may be a Heaviside function in VOF methods
or a smooth distance function in LS methods.
The local feature of the tracer function facilitates the parallelization of
Front-Capturing methods. Thus, they are generally computationally efficient.
However, a key limitation of these methods lies in their difficulty in resolving
sub-grid structures (SGS), since the interface elements are not explicitly tracked
but rather reconstructed from the tracer function.

Various algorithms have been proposed to address this limitation,
primarily within the framework of VOF methods.
The R2P method \cite{Chiodi_2020_phd, Han_2024_519}
reconstructs the interfaces within a single cell with two planes of arbitrary
relative orientation to capture a variety of SGS, such as thin films
and the closure of sheet rims.
The Moment-of-Fluid (MOF) method \cite{Dyadechko_2008_227, Jemison_2015_285, Shashkov_2023_479, Shashkov_2023_494, Chiodi_2025_528, Hergibo_2025_530} computes and evolves in time
the zeroth, first, and second moments of the fragment of material within
a computational cell. This richer geometric representation provides
more accurate reconstructions of SGS.

In Front-Tracking methods, the interface or ``front'' is represented by a set of
Lagrangian markers and their connectivity. These markers may be connected with
straight line segments \cite{Unverdi_1992_100} or global splines \cite{Popinet_1999_30},
and are advected with a given velocity field. Their connectivity must be updated
when topology changes occur, such as coalescence or breakup.
An introduction to the most popular Front-Tracking methods
can be found in \cite{Tryggvason_2011_book}. These methods provide a direct and
accurate approach for predicting SGS dynamics, but the global connectivity
information that is used to represent the interface makes parallel computing
much more challenging when compared to Front-Capturing methods, in particular
in the presence of topology changes.

In the context of multiscale simulations, geometrical properties such as skeletons
\cite{chen2022characterizing} are essential for coupling to boundary layer models
and are most naturally represented by Front-Tracking methods. 
These methods also allow for a straightforward distinction between slender objects
with a similar volume fraction distribution, including unbroken thin ligaments,
strings of small particles, and broken ligaments. This distinction is of great
relevance when analyzing statistically highly-fragmented flows \cite{Chirco_2022_467}. 

Several efforts have been made to combine features of global methods,
such as Front-Tracking, with those of local methods, such as VOF or LS.
The combination of the VOF method with marker points allows a smooth representation
of interfaces, without discontinuities \cite{Aulisa_2003_188, Aulisa_2004_197}
at cell faces or tracking of SGS \cite{Lopez_2005_208}.

Hybrid approaches that combine Front-Tracking and LS methods include
the Level Contour Reconstruction Method (LCRM) \cite{Shin_2002_180, Shin_2005_203, Shin_2007_21}, developed for structured meshes,
and the hybrid LEvel set/fronNT method (LENT) \cite{Maric_2015_113},
designed for unstructured meshes.
These methods improve the mass conservation of traditional LS methods while
explicitly avoiding storing the connectivity of Lagrangian elements. 
The idea of implicit connectivity also inspired the development of a novel 
Front-tracking method, the Local Front Reconstruction Method (LFRM) \cite{Shin_2011_230}.

We have recently presented a similar method, which is based on a purely kinematic
approach, the Edge-Based Interface-Tracking (EBIT) method \cite{Chirco_2022_95, Pan_2024_508}. 
In EBIT, the position of the interface is tracked by marker points located on the
edges of an Eulerian grid, and connectivity information is implicit. 
Moreover, markers in the EBIT method are bound to the Eulerian grid by a local reconstruction of the interface at every time step; therefore, the Eulerian grid and Lagrangian markers can be distributed to different processors by the same routine, allowing for automatic parallelization.

The basic idea and the split interface advection were discussed in \cite{Chirco_2022_95,
Pan_2024_508}, while the coupling algorithm between the EBIT method and the 
Navier--Stokes solver was presented in \cite{Pan_2024_508}.
In addition, several techniques were proposed to improve the accuracy of mass conservation and implement topology change mechanisms.

In both papers \cite{Chirco_2022_95,  Pan_2024_508}, a split scheme based on a first-order Euler time integration method is used for interface advection, in contrast
with traditional Front-Tracking methods, where an unsplit scheme is commonly used. 
The EBIT method with split advection can be more easily extended to three dimensions,
although it imposes some limitations. In particular, it is challenging to implement
high-order time integration methods and to maintain correspondence between
elements before and after each advection step, when redistributing boundary layer
quantities on the interface during the reconstruction step
\cite{Aboulhasanzadeh_2012_75,  Aboulhasanzadeh_2013_101}.
 Both difficulties can be overcome by using an unsplit scheme for interface advection
in the EBIT method.

Note that the EBIT method has also been extended to triangular meshes by Wang et 
al.~\cite{Wang_2025_520}, where an unsplit scheme is used for interface advection. The interface advection and marker reconstruction are based on the pre-image of each triangular cell, similar to the advection schemes used in the MOF method \cite{Dyadechko_2008_227, Shashkov_2023_479} and the Polygon Area Mapping (PAM) method \cite{Zhang_2008_227}.

In this paper, we propose an alternative unsplit scheme in the EBIT method 
to improve its accuracy and extend its capability for multiphase flow simulations.
The circle fit technique, used to improve mass conservation in the split scheme,
is retained. However, the algorithm to update the color vertex field has been
carefully modified to achieve consistent results.
Compared to Wang et al.'s method \cite{Wang_2025_520}, the proposed unsplit scheme is more closely related to the interface advection schemes of Front-Tracking methods,
to facilitate coupling with existing boundary layer models for multiscale
simulations. Finally, the unsplit EBIT method has been integrated into the free Basilisk platform \cite{Popinet_2003_190, Popinet_2009_228}.

The paper is organized as follows. Section~\ref{Numerical method} details
the unsplit scheme for interface advection. Then, the unsplit scheme is validated
through the computation of typical kinematic test cases in
Section~\ref{Numerical results and discussion},
The results obtained with the unsplit scheme and different time integration methods
are presented and compared with those calculated using the split scheme and the
VOF method.

\section{Numerical method} \label{Numerical method}
First, we give a brief overview of the split scheme of the original EBIT
method for interface advection to illustrate its limitations. Afterwards,
we describe in detail the implementation of the new EBIT method based on
an unsplit advection scheme, which includes the interface reconstruction
model and the algorithm to update the value of the Color Vertex field.

\subsection{A split advection scheme for the EBIT method}
% ----------
\begin{figure}
\begin{center}
\begin{tabular}{cc}
\includegraphics[width=0.45\textwidth]{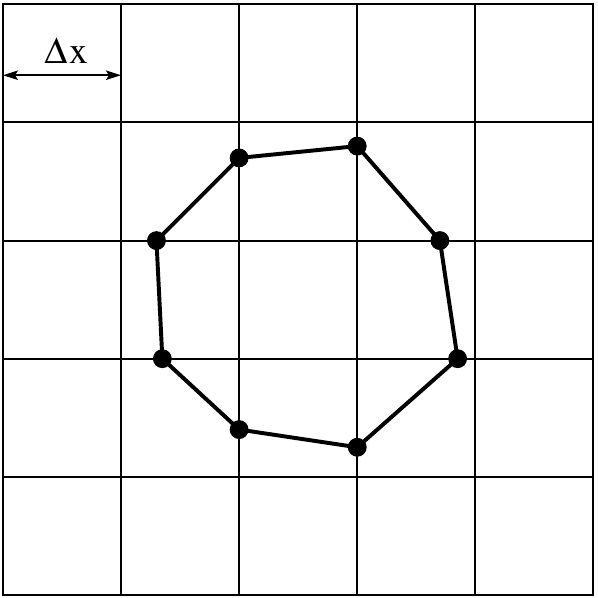} &
\includegraphics[width=0.45\textwidth]{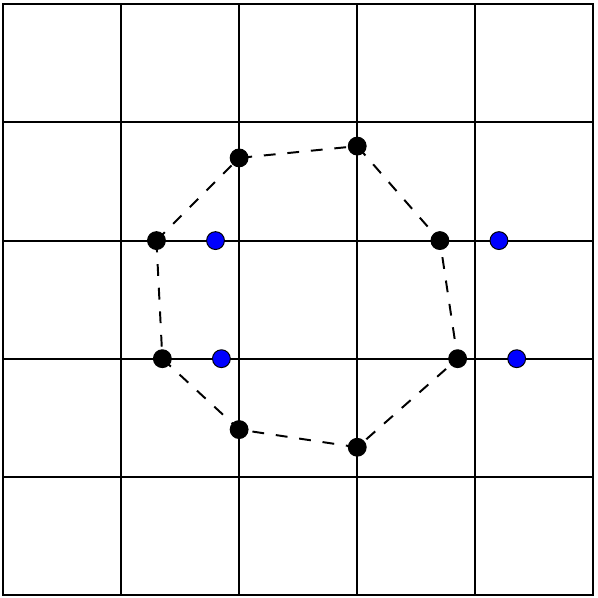}\\
(a) & (b) \\[6pt]
\includegraphics[width=0.45\textwidth]{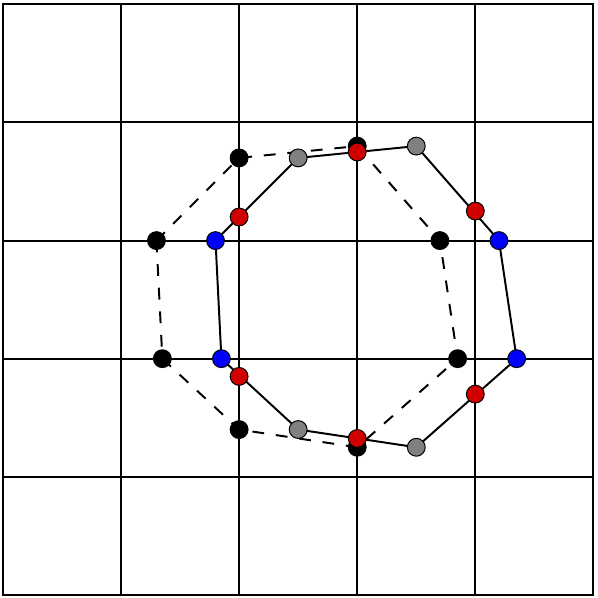} &
\includegraphics[width=0.45\textwidth]{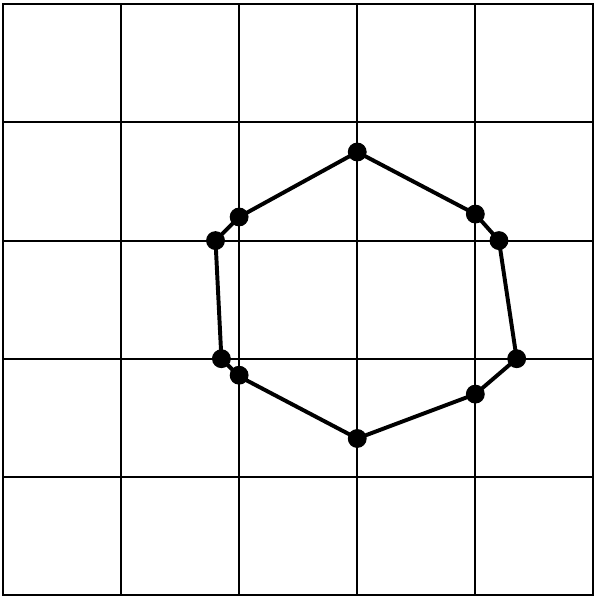}\\
(c) & (d) 
\end{tabular}
\end{center}
\caption{One-dimensional advection of the EBIT method along
the $x$-axis: (a) initial interface line; (b) advection of the markers 
on the grid lines aligned with the horizontal velocity component
(blue points); (c) advection of the unaligned markers (gray points)
and computation of the intersections with the grid lines (red points);
(d) interface line after the 1D horizontal advection}
\label{Fig_EBIT_advection_split}
\end{figure}
% ----------

In the EBIT method, the interface is represented by a set of marker points.
The markers are initially positioned on the grid lines, therefore, it is
necessary to compute the intersection points between the interface and the
grid lines at the end of each advection step. The equation of motion for a
marker point at position $\X_i$ is
% ----------
\begin{gather}
\frac{d \X_i}{dt} = \U_i \;.
\label{Eq_marker}
\end{gather}
% ----------
In our previous papers \cite{Chirco_2022_95, Pan_2024_508}, a split scheme
was used to advect the interface in a multi-dimensional computational domain,
and a first-order explicit Euler method was implemented to integrate
Eq.~\eqref{Eq_marker} in time,
% ----------
\begin{gather}
\X^{n + 1}_i= \X^{n}_i + \Delta t \, \U (\X^{n}_i, t^n) \;,
\label{Eq_marker_dis_Euler}
\end{gather}
% ----------
where $\U (\X^{n}_i, t^n)$ is the marker velocity at position $\X^{n}_i$. 
The velocity $\U$ is calculated with a bilinear interpolation of the 
discrete velocity field at time $t^n$.

\begin{figure}
\begin{center}
\begin{tabular}{cc}
\includegraphics[width=0.5\textwidth]{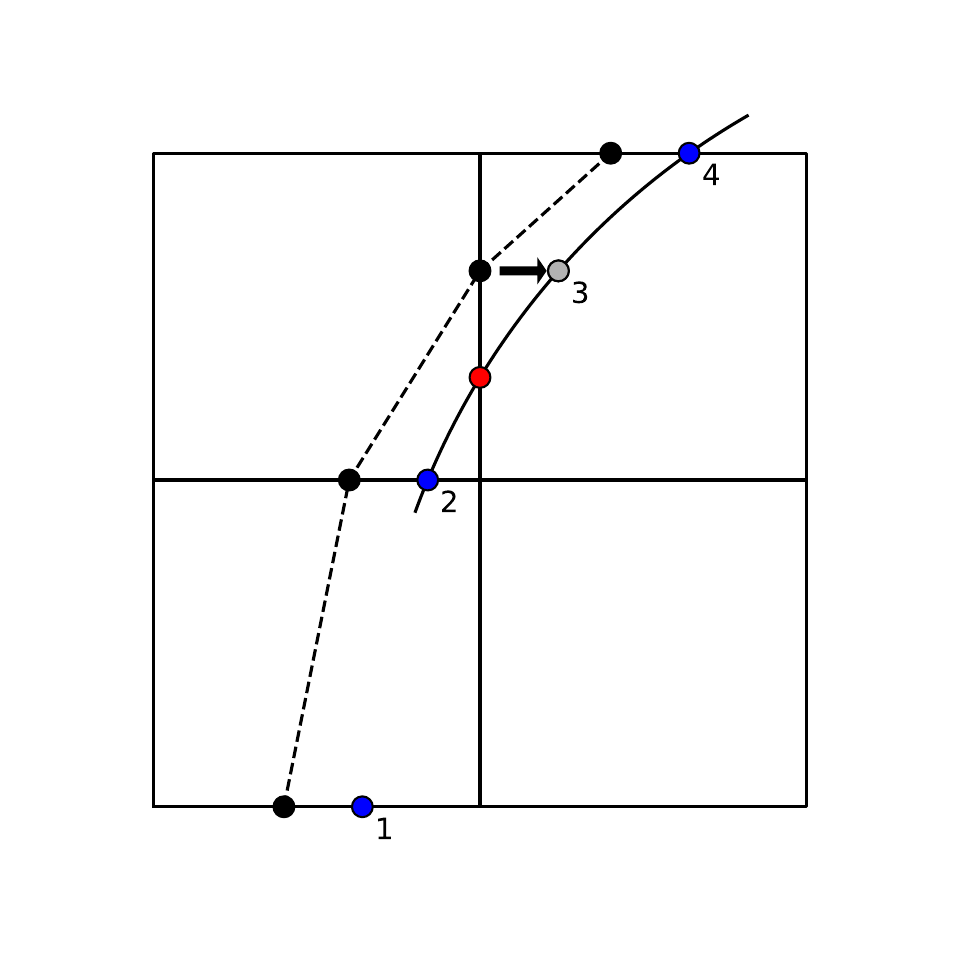} &
\includegraphics[width=0.5\textwidth]{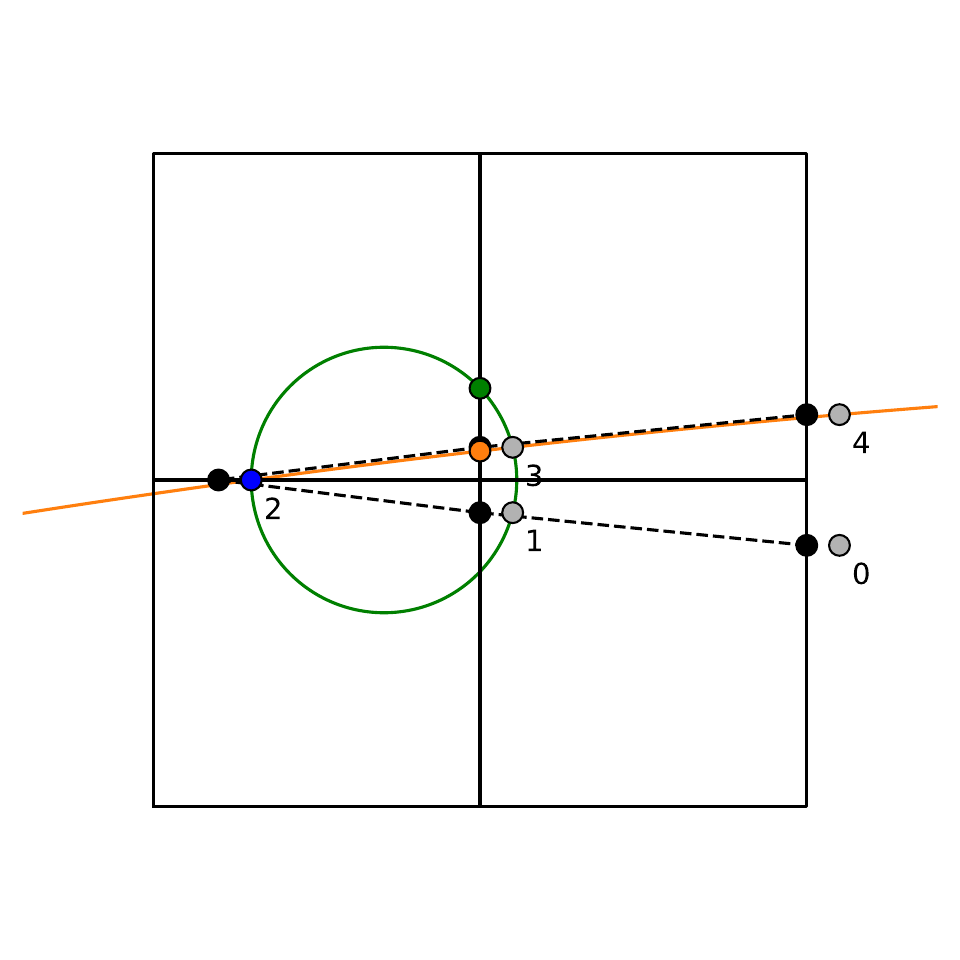}\\
(a) & (b) 
\end{tabular}
\end{center}
\caption{Positioning of the unaligned markers: (a) circle fit through 
markers $2$-$3$-$4$ to compute the intersection with the vertical grid 
line (red marker); (b) circle fits through $1$-$2$-$3$ (green line) and 
$2$-$3$-$4$ (orange line) provide two intersections far apart 
(green and orange markers)}
\label{Fig_circle_fit}
\end{figure}

In the split scheme, the marker points on the grid lines that are aligned
with the velocity component of the 1D advection are called \textit{aligned
markers}, the other ones are called \textit{unaligned markers}. 
Starting from a marker distribution at time step $n$, the new
position of the aligned markers is given by Eq.~\eqref{Eq_marker_dis_Euler}
(blue points of Fig.~\ref{Fig_EBIT_advection_split}b). To compute that 
of the unaligned markers, we first advect them again with
Eq.~\eqref{Eq_marker_dis_Euler}, to obtain the gray points of 
Fig.~\ref{Fig_EBIT_advection_split}c. Then, in the reconstruction step, 
the final position of the unaligned markers is obtained by fitting a circle
through the surrounding markers and by computing its intersection with the
corresponding grid line (red points of Fig.~\ref{Fig_EBIT_advection_split}d and of Fig.~\ref{Fig_circle_fit}a).

In the split scheme described in \cite{Pan_2024_508}, the position of the
unaligned markers was computed as the average of the results from two
different circle fits. In Fig.~\ref{Fig_circle_fit}a, the red marker
represents the intersection of the circle through markers $2$-$3$-$4$ with
the vertical grid line. Similarly, the other intersection
involves markers $1$-$2$-$3$, but it is very close to the previous one.
% ----------
% ----------

However, we have found that this simple averaging procedure can artificially
create a bulbous shape near the tip of thin ligaments, significantly
affecting the accuracy of both mass conservation and interface
representation. As shown in Fig.~\ref{Fig_circle_fit}b, the two fits yield
circles with notably different radii, near the tip region. The radius of
the green circle, through markers $1$-$2$-$3$, is much smaller than that 
of the orange circle, through markers $2$-$3$-$4$. 
However, the shape of the interface is better represented by the orange
circle with the larger radius. To address this issue, we propose an 
``ad-hoc'' approach to select the circle fit. 
Specifically, if the ratio of the two radii, $r_{max} / r_{min}$, exceeds
$10$, the circle with the largest radius is the only one used to compute
the new unaligned marker. 
The choice of the threshold value of $10$ is based on our numerical 
experiments with kinematic tests. A lower threshold may result in a less accurate reconstruction of a smooth interface, while a higher value may
hinder the algorithm's ability to detect tip structures with large
curvature variations. In general, the accuracy of mass conservation becomes
less sensitive to the specific value of this threshold as the mesh
resolution is increased.

The split scheme indeed facilitates the extension of the EBIT method to 3D,
as it can be decomposed as a sequence of 2D advection steps. However,
it also presents some disadvantages:

 i) the intermediate position of unaligned markers is lost in the
reconstruction step at the end of the 1D advection along one direction 
(gray points of Fig.~\ref{Fig_EBIT_advection_split}c). 
Moreover, in the subsequent advection along another coordinate direction,
even if we can keep track of the intermediate position of unaligned markers,
what we are actually advecting is the marker position obtained by the
reconstruction, rather than the original one. Therefore, the coupling 
of the split scheme with a high-order time integration method,
to improve the accuracy of mass conservation, as in traditional 
Front-Tracking methods \cite{Tryggvason_2011_book}, becomes rather
challenging;

 ii) in multiscale computations with Front-Tracking methods
 \cite{Aboulhasanzadeh_2012_75, Aboulhasanzadeh_2013_101}, boundary layer
 quantities, such as a scalar concentration, are typically bound to
 Lagrangian elements. Maintaining a correspondence among the elements,
 before and after the advection, is necessary to redistribute these 
 quantities during the reconstruction step in the EBIT method. However,
 it is difficult to keep track of that correspondence due to the 
 above-mentioned reasons, therefore, the EBIT method based on a split 
 scheme is less practical and attractive in multiscale simulations.

\subsection{An unsplit advection scheme for the EBIT method}

% ----------
\begin{figure}
\begin{center}
\begin{tabular}{ccc}
\includegraphics[width=0.33\textwidth]{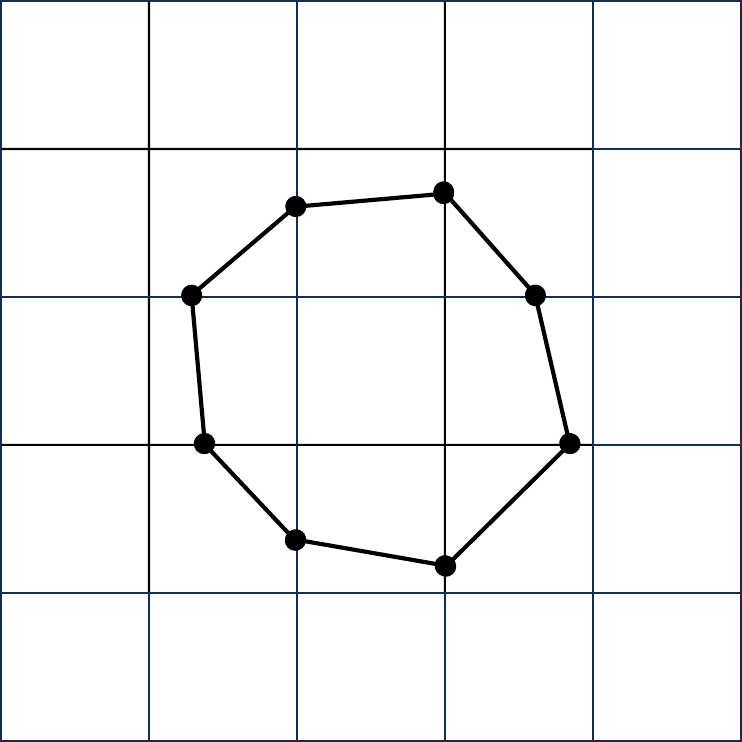} &
\includegraphics[width=0.33\textwidth]{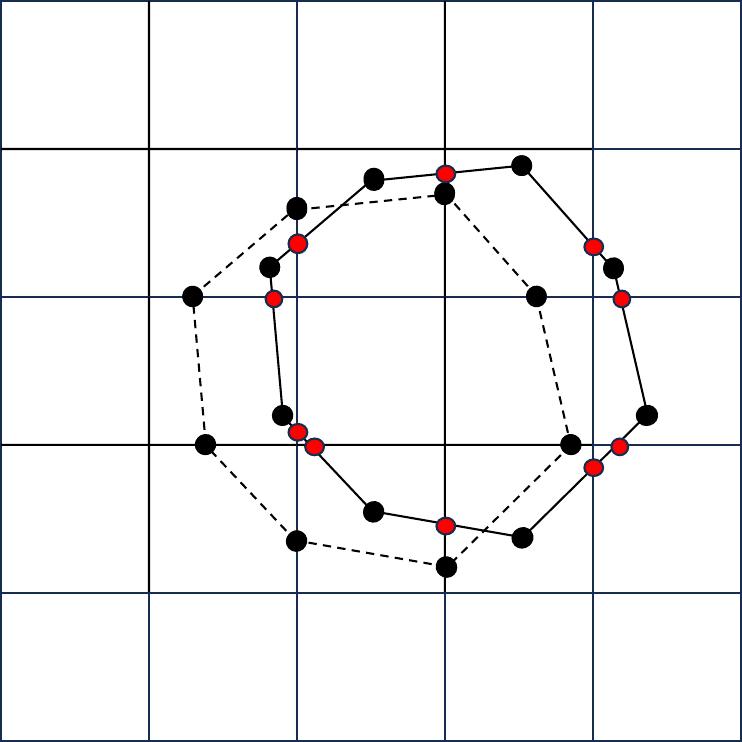} &
\includegraphics[width=0.33\textwidth]{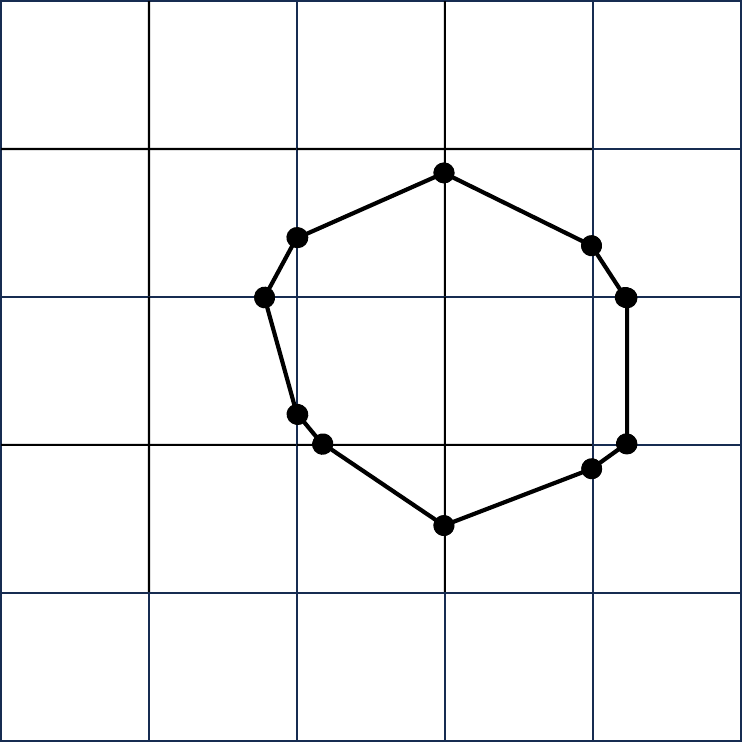}\\
(a) & (b) & (c)\\[6pt]
\end{tabular}
\end{center}
\caption{Multidimensional unsplit advection of the EBIT method: 
(a) initial interface line; (b) markers advection (black points) and
computation of the intersections with the grid lines (red points); 
(c) interface line after one step of advection}
\label{Fig_EBIT_advection_unsplit}
\end{figure}
% ----------

In this study, we consider a couple of unsplit time-integration methods 
for interface advection to extend the capability of the EBIT method. 
The advection of markers is similar to that used in traditional 
Front-Tracking methods \cite{Unverdi_1992_100}, where their final position
is predicted by integrating Eq.~\eqref{Eq_marker} with a high-order 
method (black points of Fig.~\ref{Fig_EBIT_advection_unsplit}b), but 
an additional reconstruction step, similar to that of the split scheme, 
is required at the end of each advection step (red points of 
Fig.~\ref{Fig_EBIT_advection_unsplit}b).

The circle interpolation in the reconstruction step of the split scheme
significantly improved the accuracy of mass conservation, therefore, it is
retained in the unsplit scheme. The advection is now multidimensional,
markers cannot be divided into aligned markers and unaligned ones, and the
calculation of the intersections with the grid lines needs 
to be done in the reconstruction step along all coordinate directions.

There are primarily two families of high-order time integration methods
for the advection of markers in Front-Tracking methods.
The first family consists of multistep methods \cite{Pivello_2014_58,
de_Jesus_2015_281}, such as the Adams-Bashforth method and the Adams-Moulton
method, which make use of the information from previous timesteps to achieve
high-order accuracy. However, since the EBIT markers are computed as
intersections with the grid lines in the reconstruction step, their position
at previous timesteps is not given, then multistep methods cannot be considered with EBIT.

The other family consists of multistage Runge-Kutta methods
\cite{Shin_2011_230, Terashima_2009_228, Gorges_2023_491}, where high-order
accuracy is attained by computing some intermediate values at each timestep,
instead of relying on the information from previous timesteps. 
The simplest Predictor-Corrector (PC) method \cite{Tryggvason_2011_book},
also known as Heun's method, belongs to the Runge-Kutta methods. 
In this study, the PC method and the classical fourth-order Runge-Kutta 
(RK4) will be considered for the time integration in the unsplit scheme.
The discrete form of the equation of motion \eqref{Eq_marker} for the PC
method can be written as
% ----------
\begin{gather}
\begin{cases}
\X^{n + 1}_i= \X^{n}_i + \frac{\Delta t}{2} \left(\K_1 + \K_2 \right) \;,\\
\K_1 = \U (\X^{n}_i, t^n) \;,\\
\K_2 = \U (\X^{n}_i + \Delta t \,\K_1, t^n + \Delta t) \;,
\end{cases}.
\label{Eq_marker_dis_PC}
\end{gather}
% ----------
and for the RK4 method as
% ----------
\begin{gather}
\begin{cases}
      \X^{n + 1}_i= \X^{n}_i + \frac{\Delta t}{6} 
      \left( \K_1 + 2 \K_2 + 2 \K_3  + \K_4 \right) \;,\\
\K_1 = \U (\X^{n}_i, t^n) \;,\\
\K_2 = \U (\X^{n}_i + \frac{\Delta t}{2} \,\K_1 , t^n + \frac{\Delta t}{2}) \;,\\
\K_3 = \U (\X^{n}_i + \frac{\Delta t}{2} \,\K_2 , t^n + \frac{\Delta t}{2}) \;,\\
\K_4 = \U (\X^{n}_i + \Delta t \,\K_3 , t^n + \Delta t) \;.
\end{cases}.
\label{Eq_marker_dis_RK4}
\end{gather}
% ----------

\subsection{Color vertex field} 
\label{Numerical method_color_vertex}

% ----------
\begin{figure}
\centering
\includegraphics[width=\textwidth]{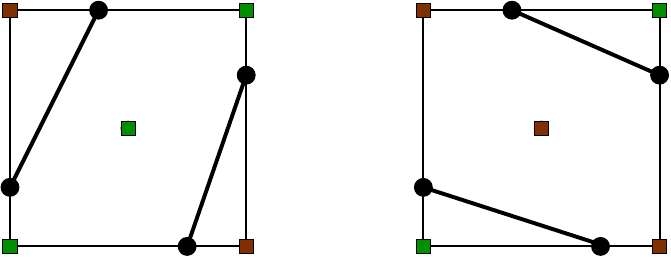}
\caption{Two different color vertex configurations, represented by 
different distributions of brown and green squares, select a different
connectivity with the same set of markers}
\label{Fig_color_vertex}
\end{figure}
% ----------

In the EBIT method, the connectivity of the markers is implicitly 
represented by the color vertex field \cite{Pan_2024_508}, which is 
a binary field in which each value is associated to one of the two fluid
phases and that locates the fluid phases in the corresponding regions
within the cell. The color vertex field is mainly used to distinguish
ambiguous configurations that are characterized by four markers present 
at the same time on the boundary of a cell, as shown in
Fig.~\ref{Fig_color_vertex}.
In other words, a one-to-one correspondence between a topological
configuration and a color vertex distribution is established within each
cell, so that the reconstruction of the interface segments can be done
without any ambiguity. The local nature of the color vertex field makes 
the EBIT method more suitable for parallelization, when compared to the
complex data structure that is used to store the connectivity in traditional
Front-Tracking methods \cite{Tryggvason_2011_book}.

% ----------
\begin{figure}
\begin{center}
\begin{tabular}{c}
\includegraphics[width=\textwidth]{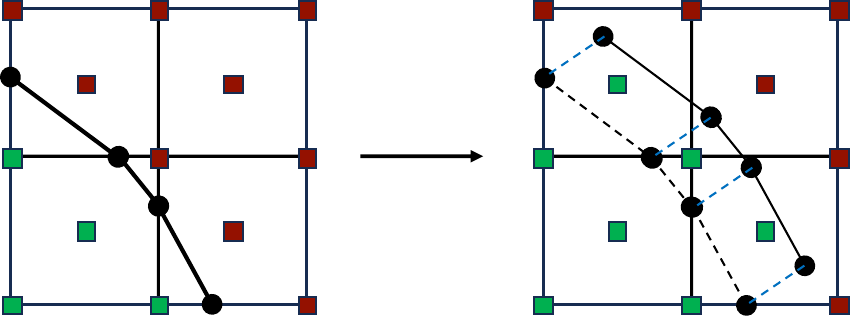}\\
(a) \\[6pt]
\includegraphics[width=0.5 \textwidth]{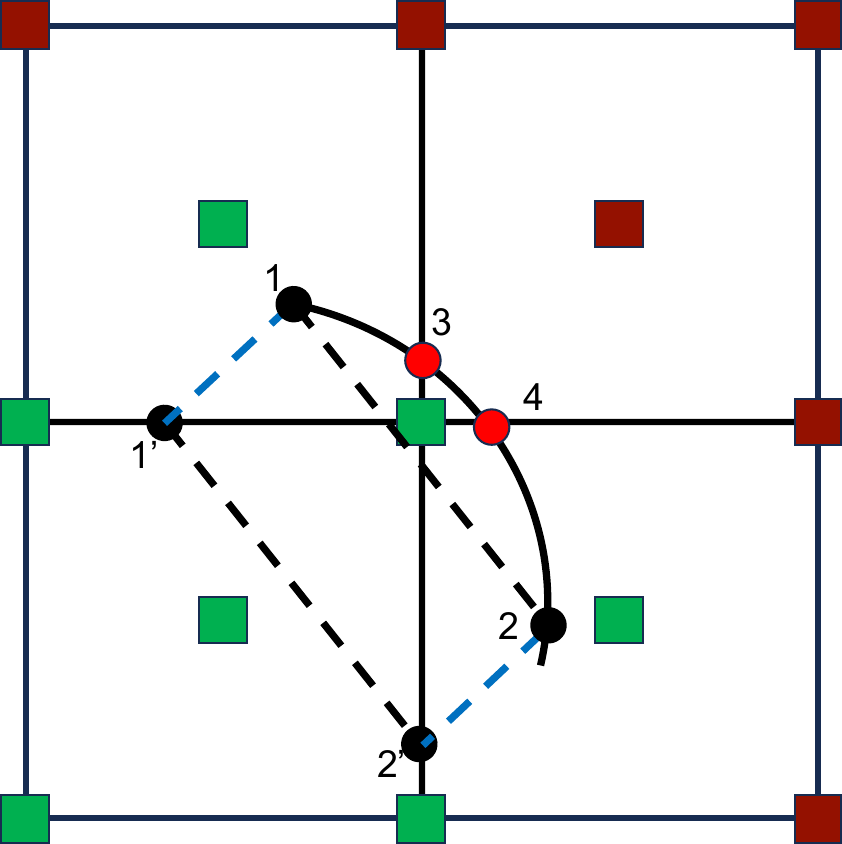}\\
(b)
\end{tabular}
\end{center}
\caption{Updating the color vertex field: (a) the color vertex value changes
in a cell corner and two cell centers, from brown to green, as each of them
is located inside a corresponding polygon formed by connecting the four
endpoints of an interface segment before and after its advection; 
(b) the intersection points (red points) should also be considered in the
construction of the polygon to obtain a consistent result with
the circle interpolation}
\label{Fig_color_vertex_update}
\end{figure}
% ----------

As the interface is advected, the color vertex field should also be updated
accordingly, to ensure that the implicit connectivity information is
retained. For the split scheme, a detailed description of how to update the
color vertex value on a cell corner and on the cell center was provided 
in \cite{Pan_2024_508}.

For an unsplit scheme, the algorithm is simpler: we change the value
of a color vertex if an interface segment sweeps its position. 
The same algorithm can be applied to update the color vertex value 
on the four corners and the center of a computational cell. 
To determine if a point of the computational domain is swept by the 
advection of an interface segment, we test whether its position is located
inside the polygon formed by connecting the four endpoints of an interface
segment before and after its advection, as shown in
Fig.~\ref{Fig_color_vertex_update}a. 

To perform this test, we use the robust ``ray casting'' algorithm developed
in the field of computational geometry to perform the point-in-polygon 
(PIP) test.
A ray is traced from the point under consideration along a fixed direction,
and the number of intersections between this ray and the polygon edges is
computed. This integer number will be an even number if the point is outside
the polygon, or an odd number if the point is inside the polygon.
In our implementation, the ray is directed along the positive  horizontal
direction, but with a very small intersection angle $\theta$ with the 
$x$-axis, $\theta = 10^{-32}$, to avoid the difficulty associated 
with corner cases, where the ray passes exactly through a vertex of the polygon.

Since the position of the intersections with the grid lines is computed
with the circle interpolation at the end of an advection step, see the red
points of Fig.~\ref{Fig_color_vertex_update}b, the construction of
a polygon with only four endpoints may give rise to inconsistent results when
,after its advection, the interface segment is very close to a cell corner. 
In Fig.~\ref{Fig_color_vertex_update}b, the color vertex value in the
middle of the figure should change from brown to green, but the cell corner
is outside the polygon $1'$-$2'$-$2$-$1$. In this case, we have
also to consider the red points of Fig.~\ref{Fig_color_vertex_update}b, 
to construct the polygon $1'$-$2'$-$2$-$4$-$3$-$1$ with six edges, 
to obtain a consistent result.

\section{Numerical results and discussion} \label{Numerical results and discussion}
\subsection{Translation with uniform velocity}

In this test, a circle of radius $R=0.15$ and center at $(0.25, 0.75)$ is placed inside the
unit square domain. The reference phase is always positioned inside the circle. 
The computational domain is subdivided into square cells of size
$h =1/N_x$, where $N_x = 32, 64, 128, 256, 512$. A uniform and constant velocity field
$(u, v)$, where $u = -v$, is applied, so that the circular interface is advected along a
diagonal direction. At half time $t = 0.5\,T$, the center reaches the position $(0.75, 0.25)$, 
the velocity field is then reversed, and the circle should return to its initial position
at time $t = T = 1$ without distortion.

We consider this simple velocity field mainly to test the two algorithms for the interface
reconstruction and for the evolution of the color vertex field with the unsplit scheme. 
Since a uniform velocity field is applied, and the magnitude of the two velocity components
is the same, we expect no major discrepancy between the results obtained with the split
scheme and with the unsplit one. 

The area, shape, and symmetric difference errors measure the accuracy of the method and the conservation of mass. 
The area error $E_{area}$ is defined as the absolute value of the relative difference
between the area $A(0)$ occupied by the reference phase at the initial time $t=0$ and the 
area $A(T)$ at time $t=T$
% ----------
\begin{equation}
E_{area} = \frac{|A(T) - A(0)|}{A(0)} \;.
\label{Eq_error_surface}
\end{equation}
% ----------

The shape error $E_{shape}$, in $L_\infty$ norm, is defined as the maximum distance between
any marker $\boldsymbol{x}_i$ on the interface and the corresponding closest point on the analytical solution. For a circular interface, the shape error is simply the following
% ----------
\begin{equation}
E_{shape} = \max\limits_i |\textrm{dist} (\boldsymbol{x}_i)|, \quad
\textrm{dist} (\boldsymbol{x}_i) = \sqrt{(x_i - x_c)^2 + (y_i - y_c)^2} - R \;,
\label{Eq_error_shape}
\end{equation}
% ----------
where $(x_c, y_c)$ are the coordinates of the center and $R$ the radius. For all kinematic
tests discussed in the following sections, the shape error is evaluated at the end of the
simulation.

Given the two domains $A$ and $B$, the symmetric difference $A \bigtriangleup B$ is 
defined as 
% ----------
\begin{equation}
A \bigtriangleup B = (A \cup B) \,\backslash\, (A \cap B) \;.
\label{Eq_sym_diff}
\end{equation}
% ----------
\noindent
In the tests presented in this section, the interface reconstruction at the beginning
of the simulation is $A$ and that at the end of the simulation is $B$. We use the area
$E_{sym}$ of the symmetric difference,
% ----------
\begin{equation}
E_{sym} = |A \bigtriangleup B| = |A| + |B| - 2 |A \cap B| \;,
\label{Eq_sym_diff_error}
\end{equation}
% ----------
to measure the accuracy of an interface advection scheme.

% ----------
\begin{figure}
\begin{center}
\begin{tabular}{cc}
\includegraphics[width=0.45\textwidth]{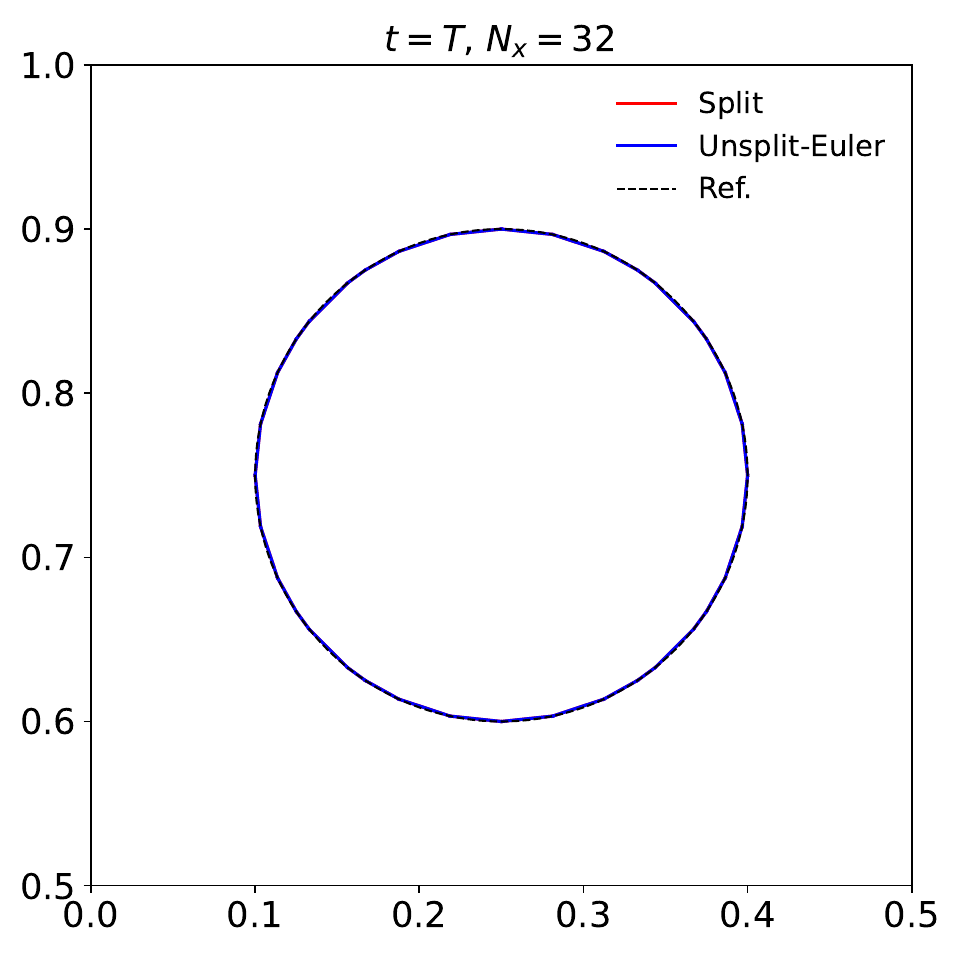} &
\includegraphics[width=0.45\textwidth]{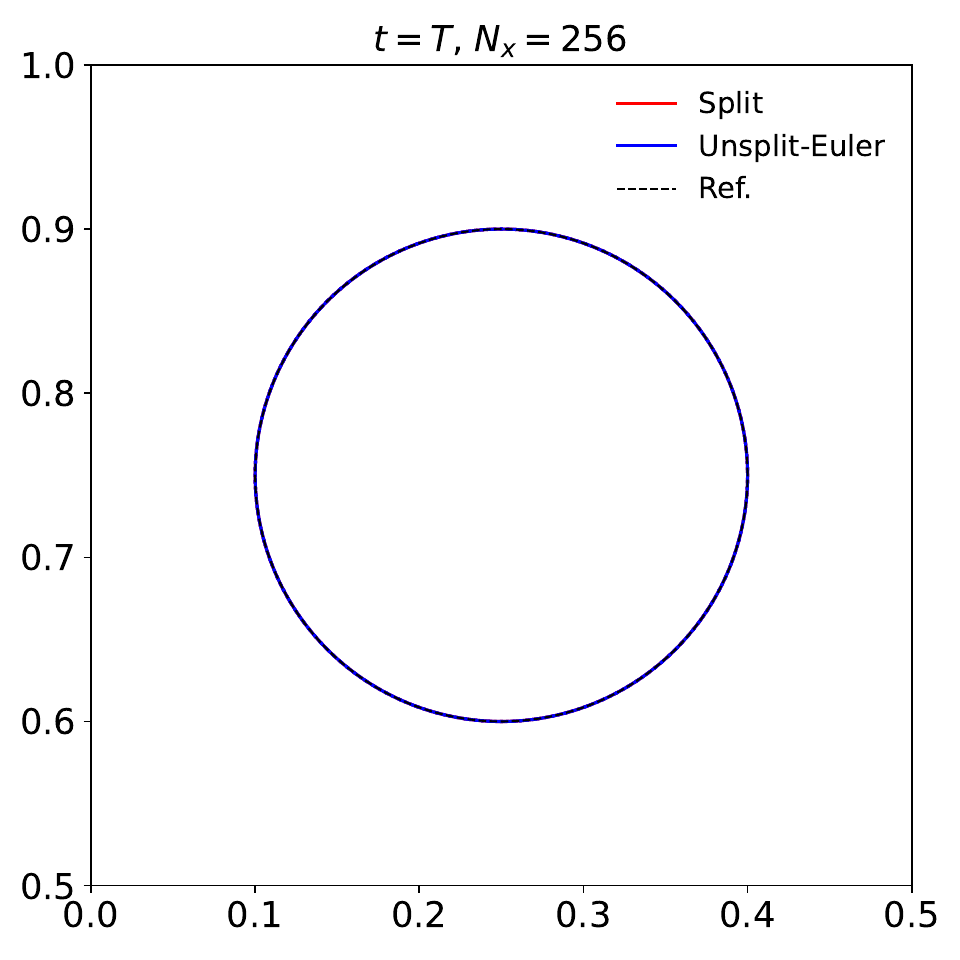}\\
(a) & (b) 
%\\
%\includegraphics[width=0.45\textwidth]{translation_intf_128} &
%\includegraphics[width=0.45\textwidth]{translation_intf_256}\\
%(c) & (d) 
\end{tabular}
\end{center}
\caption{Interface lines at the end of the translation test for different grid 
resolutions: (a) $N_x = 32$; (b) $N_x = 256$
%; (c) $N_x = 128$; (d) $N_x = 256$
}
\label{Fig_translation_intf}
\end{figure}
% ----------

% ----------
\begin{table}[hbt!]
\footnotesize
\caption{Mesh convergence study for the translation test}
\centering
\begin{tabular}{cc|ccccc}
\hline 
 &$N_x$& 32 & 64 & 128 & 256 & 512 \\ 
\hline 
Split &$E_{area}$ & $9.32\times 10^{-9}$ & $2.30 \times 10^{-9}$ & $3.13 \times 10^{-10}$ & $1.48 \times 10^{-10}$ & $3.82 \times 10^{-11}$ \\ 
&$E_{shape}$ & $6.13 \times 10^{-9}$ & $3.21 \times 10^{-9}$ & $3.77 \times 10^{-9}$ & $9.08 \times 10^{-10}$ & $3.76 \times 10^{-10}$ \\
&$E_{sym}$ &$2.71 \times 10^{-9}$ & $2.81 \times 10^{-9}$ & $1.37 \times 10^{-9}$ & $5.09 \times 10^{-10}$ & $2.28 \times 10^{-10}$ \\
\hline
Unsplit-Euler &$E_{area}$ & $8.03 \times 10^{-9}$ & $2.32 \times 10^{-9}$ & $9.83 \times 10^{-10}$ & $1.14 \times 10^{-10}$ & $2.68 \times 10^{-11}$\\
&$E_{shape}$ & $5.22 \times 10^{-9}$ & $2.76 \times 10^{-9}$ & $3.57 \times 10^{-9}$ & $8.14 \times 10^{-10}$ & $3.75 \times 10^{-10}$ \\
&$E_{sym}$ & $2.70 \times 10^{-9}$ & $2.82 \times 10^{-9}$ & $1.20 \times 10^{-9}$ & $5.12 \times 10^{-10}$ & $2.29 \times 10^{-10}$ \\
\hline 
\end{tabular}
\label{Tab_translation_error}
\normalsize
\end{table}
% ----------
% ----------
\begin{figure}
\begin{center}
\begin{tabular}{ccc}
\includegraphics[width=0.33\textwidth]{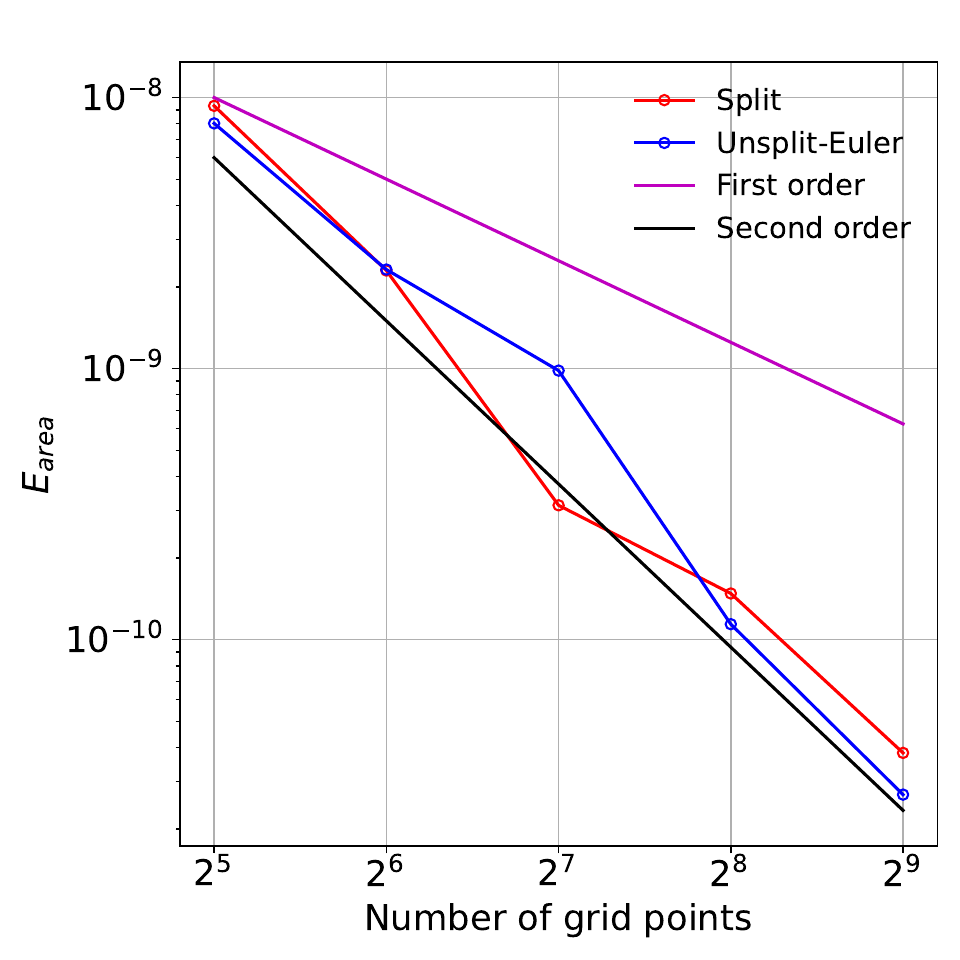} &
\includegraphics[width=0.33\textwidth]{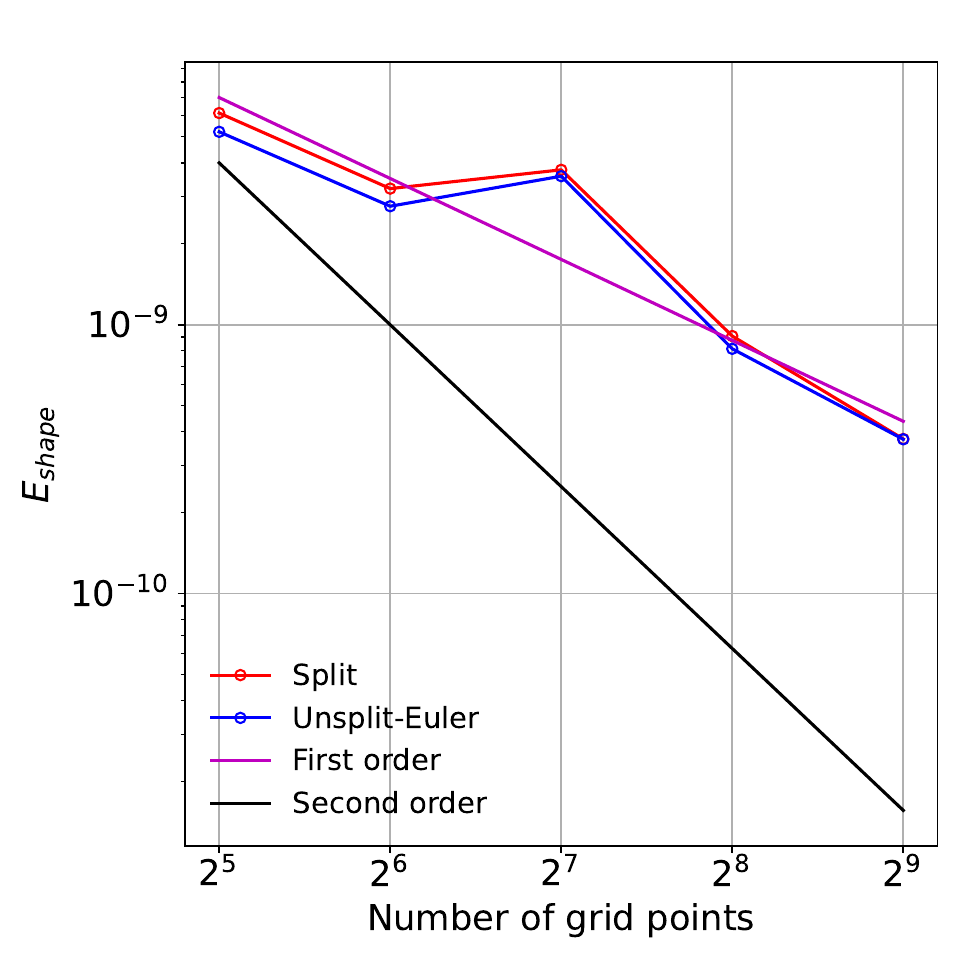} &
\includegraphics[width=0.33\textwidth]{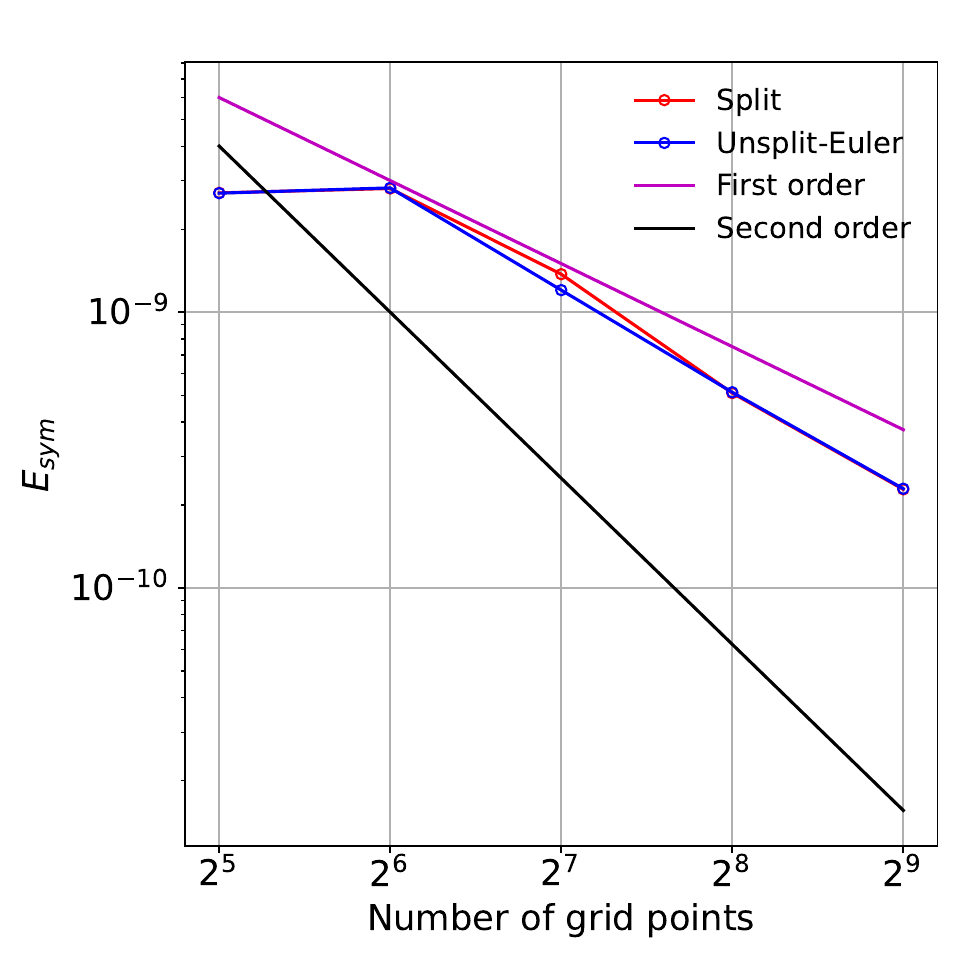}\\
(a) & (b) & (c)
\end{tabular}
\end{center}
\caption{Errors in the translation test as a function of grid resolution: (a) area error $E_{area}$; (b) shape error $E_{shape}$; (c) symmetric difference error $E_{sym}$}
\label{Fig_translation_error}
\end{figure}
% ----------

To initialize the markers on the grid lines, we first compute the signed distance
\eqref{Eq_error_shape} on the cell vertices; then we use a root-finding routine
to calculate the position of a marker when the sign of the distance is opposite
on the two endpoints of a cell side. There is a small numerical error in the initial data,
due to the tolerance of the root-finding routine, that accumulates as the interface is
translated. However, because of the circle fit in the EBIT method, this error remains
rather limited during translation.

We consider a relatively small $\textrm{CFL}$ number
$\textrm{CFL} = (u\,\Delta t)/ h = 0.125$. 
Since the velocity field changes discontinuously at half time,
only the first-order explicit Euler method is used for the time integration for both
the split and unsplit schemes. The corresponding results are denoted by ``Split'' and
``Unsplit-Euler'' in the table and figure. The interface lines at the end of the
simulation are shown in Fig.~\ref{Fig_translation_intf}, for the coarsest and most refined
grid resolutions. The interface lines of the split and unsplit schemes overlap very well,
as it was expected, since the velocity field is uniform and a circle fit is used
in the reconstruction step.

% discussion about the result on N = 128?
The area error $E_{area}$, the shape error $E_{shape}$ and the symmetric difference
error $E_{sym}$ are listed in Table~\ref{Tab_translation_error} and are shown in
Fig.~\ref{Fig_translation_error}.
All the errors obtained with the two advection schemes are in good agreement with each
other, except the area error at mesh resolution $N_x = 128$. However, the magnitude of the
difference is tiny, $\Delta E_{area} \approx 6 \times 10^{-10}$, given that
the area error is already quite small. 
Second-order convergence is observed for both schemes for the area error, while
only first-order convergence is found for the shape error and the symmetric difference
error. The strong agreement between the results from different advection schemes
also validates the algorithm for updating the color vertex field described in
Section~\ref{Numerical method_color_vertex}.

\subsection{Solid body rotation}

A circular interface of radius $R = 0.15$ and center at $(0.5, 0.75)$ is placed
again inside the unit square domain. A constant velocity field $\U$ is applied
throughout the region,
$\U=(u, v) = (2\pi (0.5 - y), 2\pi (x - 0.5))$, so that the interface
rotates around the center of the computational domain and returns to its initial
position at time $t = T = 1$. 
Each velocity component is a linear function of one Cartesian coordinate, and the
bilinear interpolation method does not introduce any numerical approximation in the
computation of a marker velocity. Therefore, the numerical errors are only due to the
reconstruction step and the time integration method.

In the simulations, we use a constant timestep that varies with the mesh resolution. 
Its value is determined by the maximum horizontal component of the velocity, so that
$\textrm{CFL} = (u_{max}\, \Delta t) / h = \pi / 16 \approx 0.2$. The accuracy
of the advection schemes is again measured by the area, shape, and symmetric
difference errors.

% ----------
\begin{figure}
\begin{center}
\begin{tabular}{cc}
\includegraphics[width=0.45\textwidth]{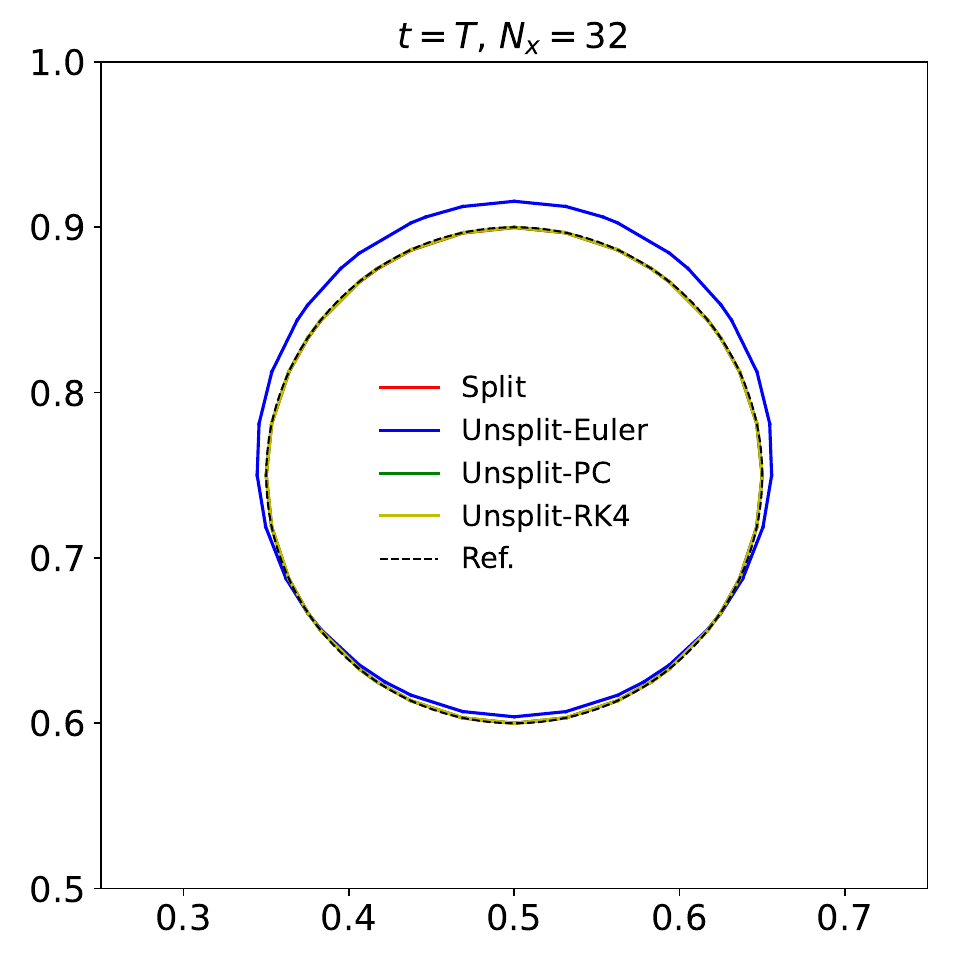} &
\includegraphics[width=0.45\textwidth]{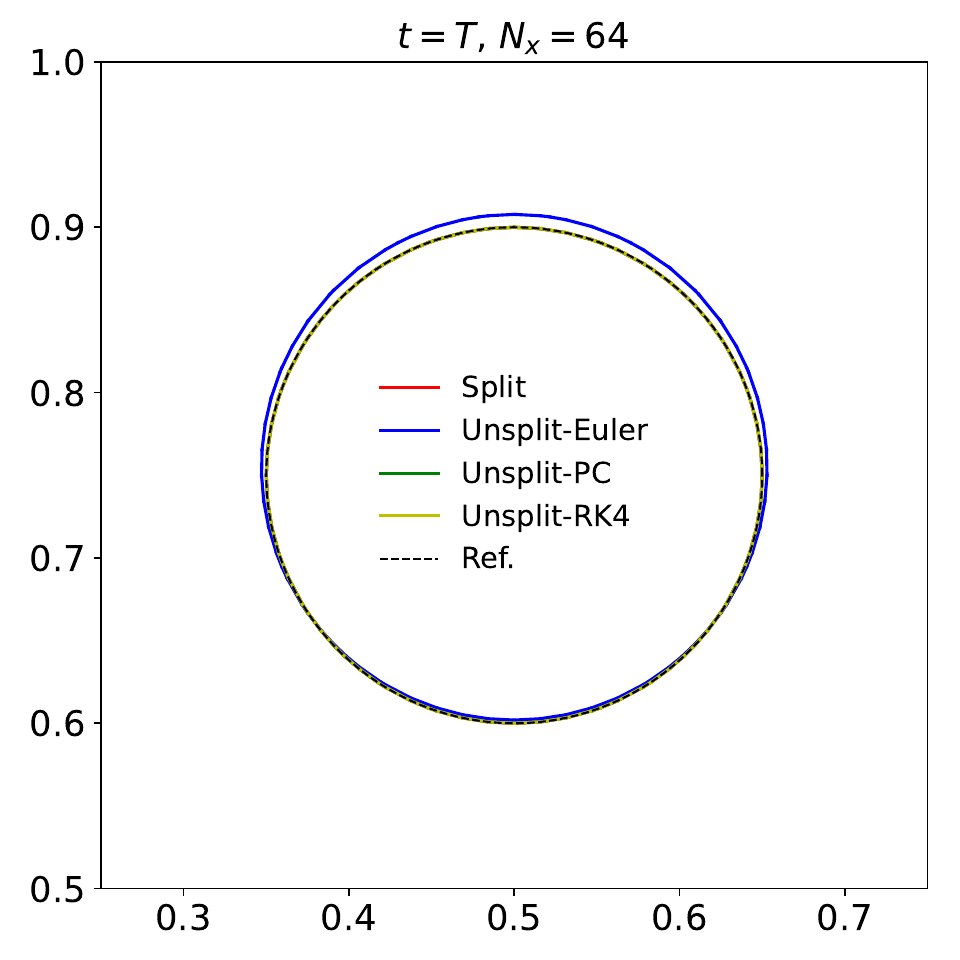}\\
(a) & (b) \\
\includegraphics[width=0.45\textwidth]{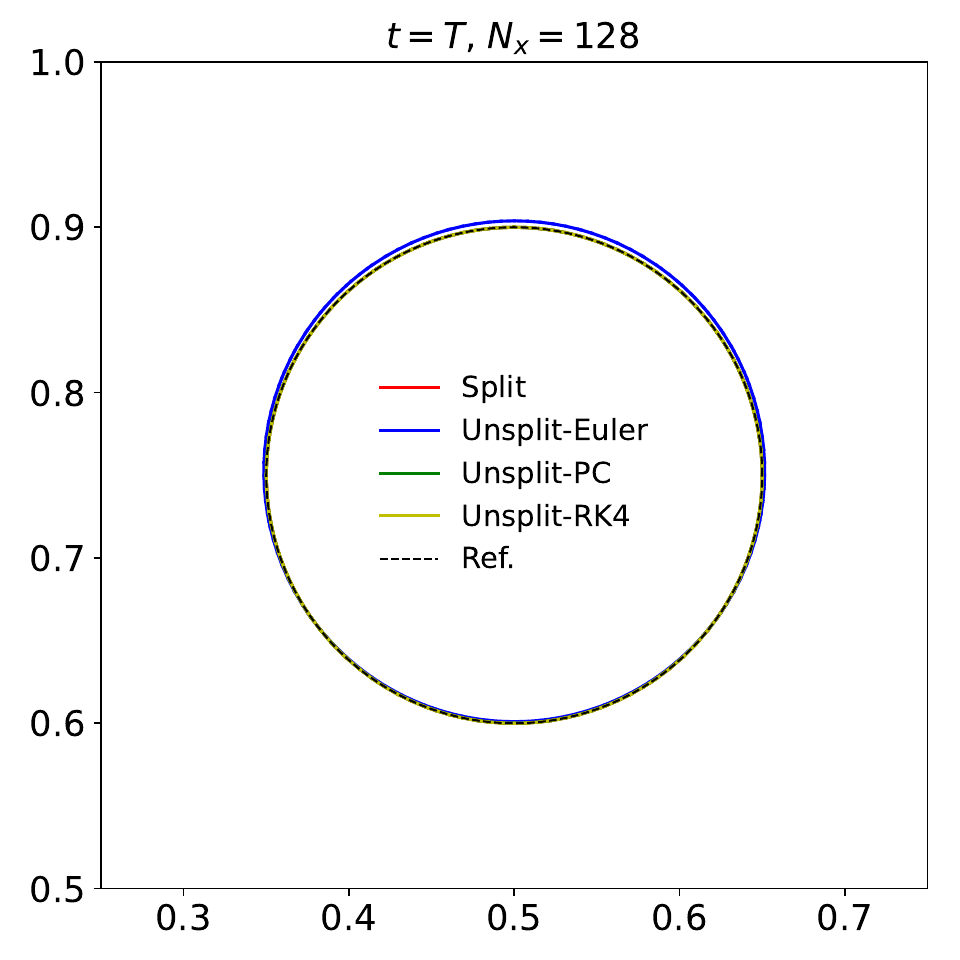} &
\includegraphics[width=0.45\textwidth]{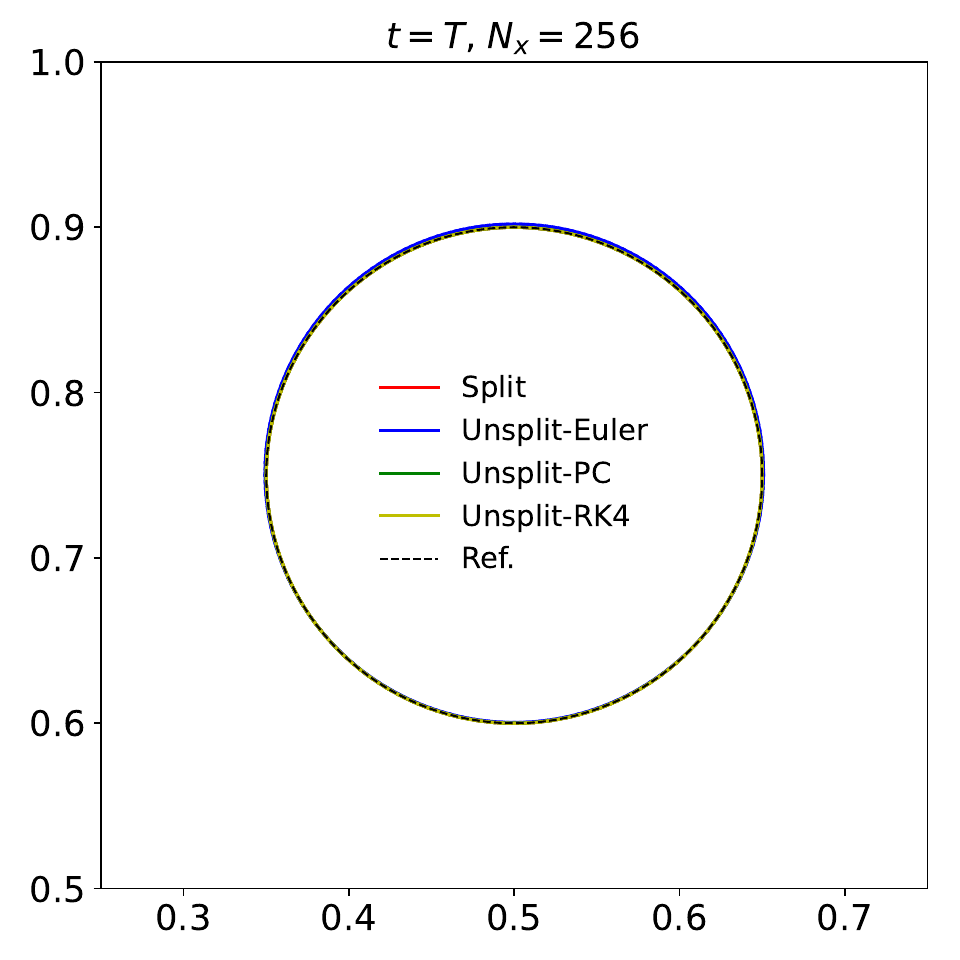}\\
(c) & (d) 
\end{tabular}
\end{center}
\caption{Interface lines at the end of the rotation test for different grid resolutions:
(a) $N_x = 32$; (b) $N_x = 64$; (c) $N_x = 128$; (d) $N_x = 256$}
\label{Fig_rotation_intf}
\end{figure}
% ----------

% ----------
\begin{table}[hbt!]
\footnotesize
\caption{Mesh convergence study for the rotation test}
\centering
\begin{tabular}{cc|ccccc}
\hline 
 &$N_x$& 32 & 64 & 128 & 256 & 512 \\ 
\hline 
Split &$E_{area}$ & $6.21\times 10^{-4}$ & $3.67 \times 10^{-5}$ & $9.00 \times 10^{-6}$ & $1.87 \times 10^{-6}$ & $1.32 \times 10^{-7}$ \\ 
&$E_{shape}$ & $2.61 \times 10^{-4}$ & $1.82 \times 10^{-5}$ & $8.28 \times 10^{-6}$ & $1.74 \times 10^{-6}$ & $3.29 \times 10^{-7}$ \\
&$E_{sym}$ &$6.07 \times 10^{-4}$ & $4.79 \times 10^{-6}$ & $1.04 \times 10^{-6}$ & $2.36 \times 10^{-7}$ & $3.11 \times 10^{-8}$ \\
\hline
Unsplit-Euler &$E_{area}$ & $8.05 \times 10^{-2}$ & $3.91 \times 10^{-2}$ & $1.95 \times 10^{-2}$ & $9.68 \times 10^{-3}$ & $4.83 \times 10^{-3}$\\
&$E_{shape}$ & $1.57 \times 10^{-2}$ & $7.78 \times 10^{-3}$ & $3.87 \times 10^{-3}$ & $1.93 \times 10^{-3}$ & $9.65 \times 10^{-4}$ \\
&$E_{sym}$ & $7.09 \times 10^{-3}$ & $3.48 \times 10^{-3}$ & $1.73 \times 10^{-3}$ & $8.61 \times 10^{-4}$ & $4.30 \times 10^{-4}$ \\
\hline 
Unsplit-PC &$E_{area}$ & $2.83 \times 10^{-6}$ & $3.63 \times 10^{-7}$ & $4.47 \times 10^{-8}$ & $5.65 \times 10^{-9}$ & $7.08 \times 10^{-10}$\\ 
&$E_{shape}$ & $3.96 \times 10^{-5}$ & $9.88 \times 10^{-6}$ & $2.46 \times 10^{-6}$ & $6.16 \times 10^{-7}$ & $1.54 \times 10^{-7}$\\
&$E_{sym}$ & $2.30 \times 10^{-5}$ & $5.95 \times 10^{-6}$ & $1.47 \times 10^{-6}$ & $3.69 \times 10^{-7}$ & $9.24 \times 10^{-8}$\\
\hline
Unsplit-RK4 &$E_{area}$ & $2.39\times 10^{-11}$ & $7.54 \times 10^{-13}$ & $8.05 \times 10^{-15}$ & $4.16 \times 10^{-14}$ & $7.60 \times 10^{-14}$\\ 
&$E_{shape}$ & $2.99 \times 10^{-10}$ & $1.86 \times 10^{-11}$ & $1.16 \times 10^{-12}$ & $7.77 \times 10^{-14}$ & $3.48 \times 10^{-14}$\\
&$E_{sym}$ & $1.73 \times 10^{-10}$ & $1.12 \times 10^{-11}$ & $6.91 \times 10^{-13}$ & $4.46 \times 10^{-14}$ & $6.91 \times 10^{-15}$\\
\hline 
\end{tabular}
\label{Tab_rotation_error}
\normalsize
\end{table}
% ----------
% ----------
\begin{figure}
\begin{center}
\begin{tabular}{ccc}
\includegraphics[width=0.33\textwidth]{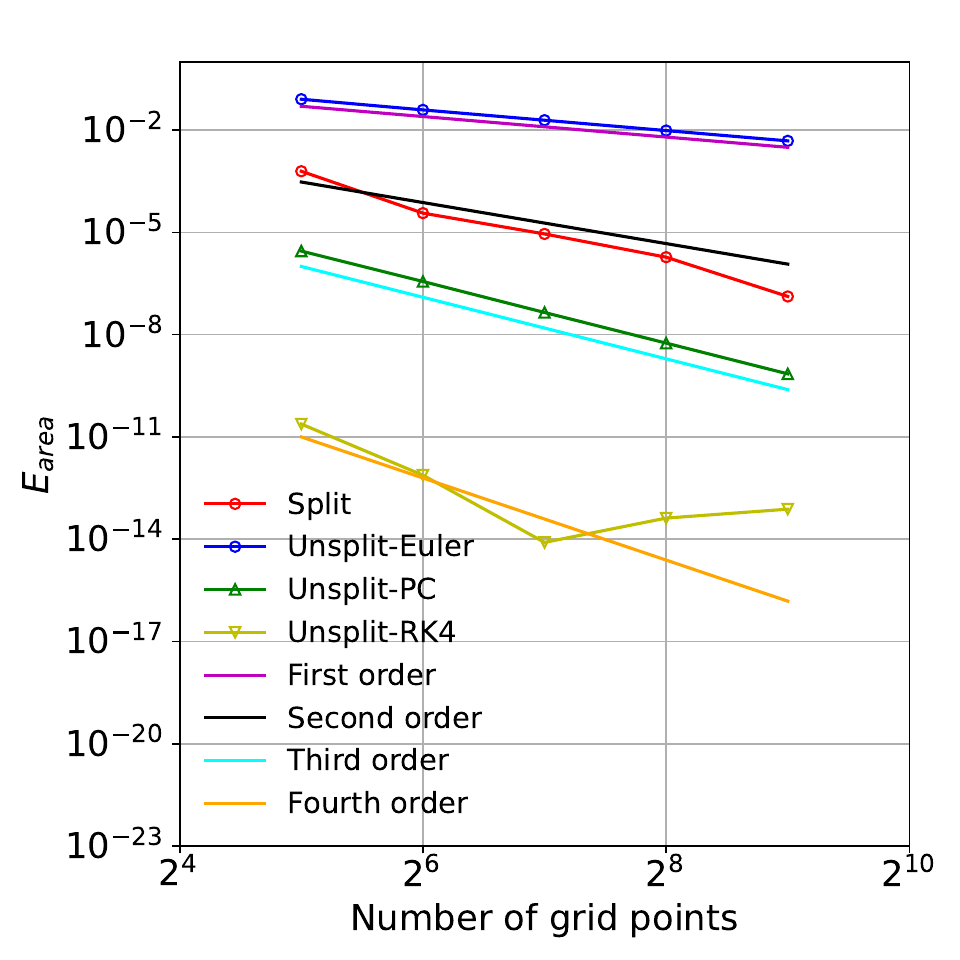} &
\includegraphics[width=0.33\textwidth]{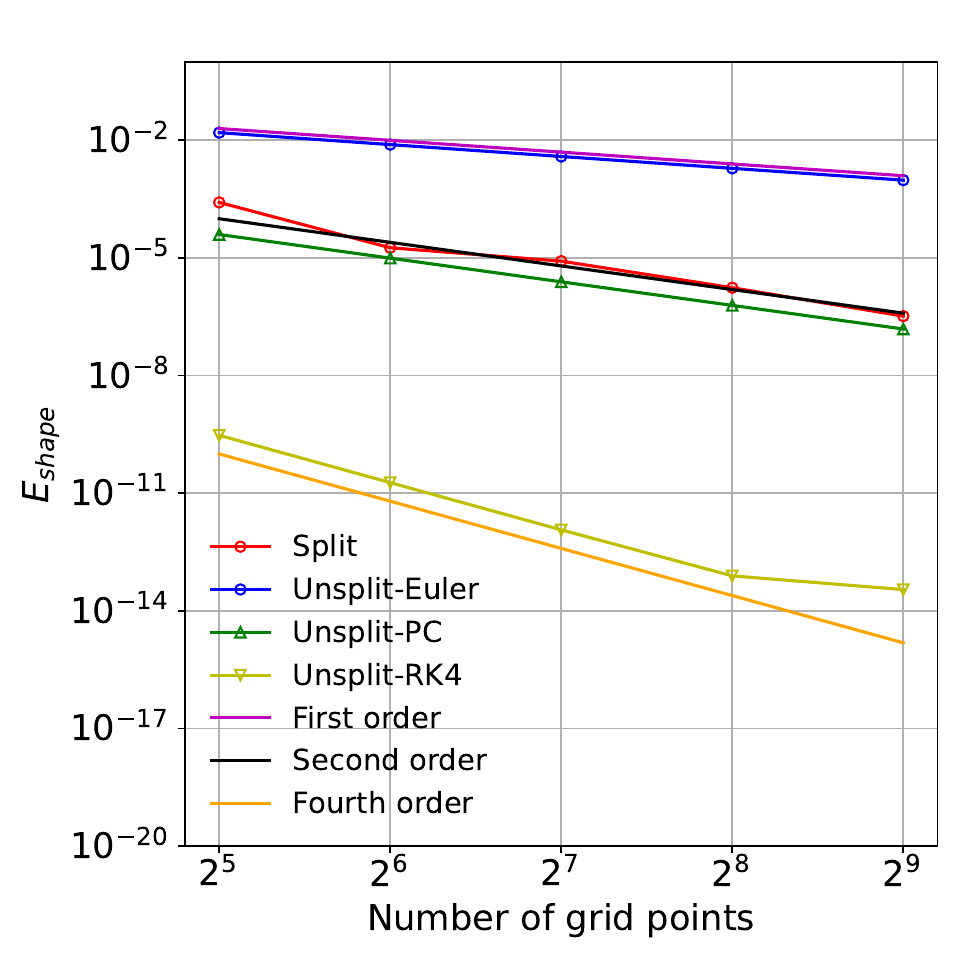} &
\includegraphics[width=0.33\textwidth]{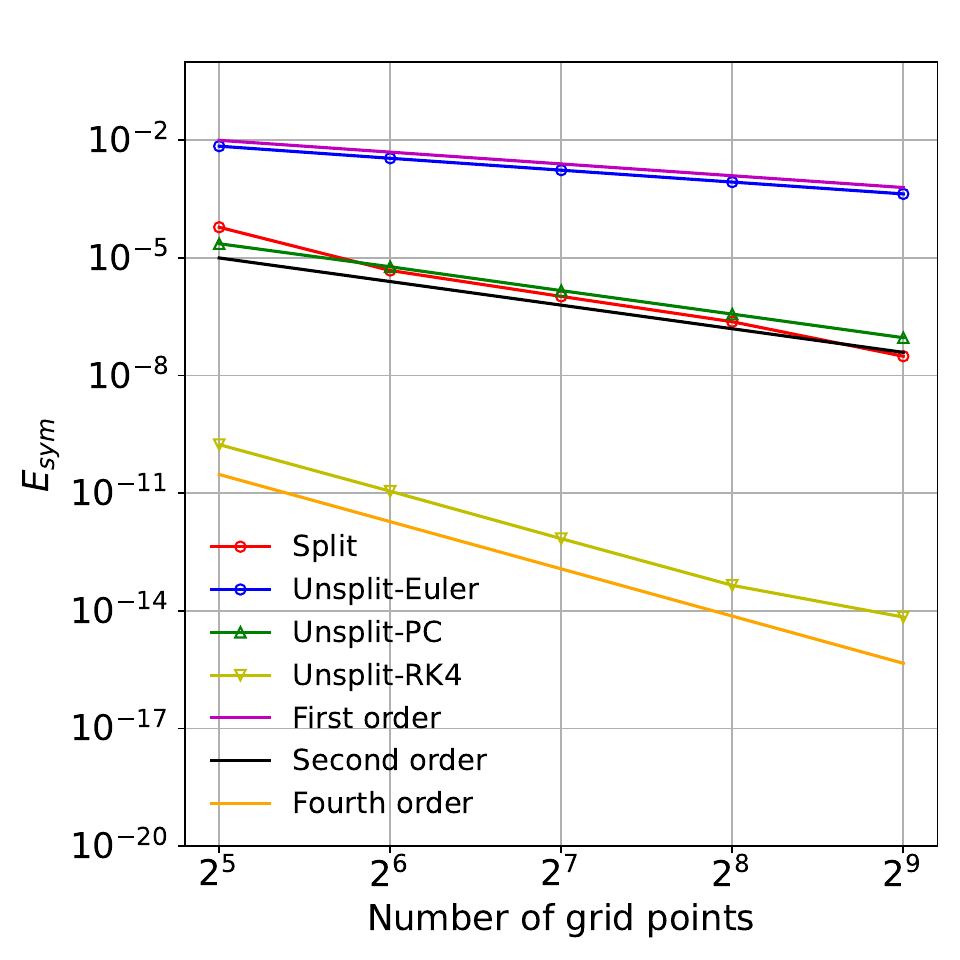}\\
(a) & (b) & (c)
\end{tabular}
\end{center}
\caption{Errors in the rotation test as a function of grid resolution: (a) area error $E_{area}$; (b) shape error $E_{shape}$; (c) symmetric difference error $E_{sym}$}
\label{Fig_rotation_error}
\end{figure}
% ----------

The interface lines at the end of the simulation are presented in Fig.~\ref{Fig_rotation_intf}
for different advection schemes and mesh resolutions. Furthermore, in this test, we also
consider two high-order time integration methods, the PC and the RK4 methods, denoted by
``Unsplit-PC'' and by ``Unsplit-RK4'', respectively.

For the interface lines obtained with the unsplit-Euler method, we observe an upward shift
and an expansion of the upper half of the interface for all mesh resolutions. The results
with the unsplit-PC, the unsplit-RK4, and the split schemes agree much better with
the reference solution.
 
It is worth noting the remarkable difference between the results of the split and
Unsplit-Euler schemes, even if the same first-order Euler method is used for the time
integration. We can understand why the split scheme is more accurate by examining
Eq.~\eqref{Eq_marker_dis_Euler} for an aligned marker along the $x$-direction. Its final position at the end of an advection step is given by
% ----------
\begin{gather}
\begin{cases}
x^{n + 1}_i= x^{n}_i + \Delta t \,k_{1, x} \;, \\
y^{n + 1}_i= y^{n}_i + \Delta t \,k_{2, y} \;, \\
k_{1, x}  = u_x (x^{n}_i, y^{n}_i, t^n) \;, \\
k_{2, y} = u_y (x^{n}_i + \Delta t \,k_{1,x}, y^{n}_i, t^n) \;.
\end{cases} \;.
\label{Eq_marker_dis_Euler_split}
\end{gather}
% ----------

The $y$-component of the marker velocity $u_y$ is evaluated at the intermediate position
$(x^{n}_i + \Delta t\, k_{1,x}, y^{n}_i)$, which can be viewed as a correction step, similar
to that of the PC method, even if here the correction is at time $t^n$. Therefore, we expect
that a split scheme is more accurate than an unsplit scheme when both of them are
coupled with a first-order explicit Euler method.

The area error $E_{area}$, the shape error $E_{shape}$ and the symmetric difference error
$E_{sym}$ are listed in Table~\ref{Tab_rotation_error} and are shown in
Fig.~\ref{Fig_rotation_error}.
For the unsplit-Euler method, only first-order convergence is observed for all errors. 
For the unsplit-PC method, third-order convergence is observed for the area error, while
second-order convergence is observed for the shape and symmetric difference errors.
For the unsplit-RK4 method, fourth-order convergence is observed for all errors; however, 
as the mesh resolution is increased, the area error soon reaches the machine zero level and
starts to increase slightly at the two finest resolutions ($N_x = 256, 512$).

For the split scheme, the three errors are much smaller than those obtained with the 
unsplit-Euler method, and second-order convergence is observed for all of them. 
Furthermore, the shape and symmetric difference errors are close to those obtained with the
unsplit-PC method. By comparing the discretized equations of motion of these two methods, 
Eqs.~\eqref{Eq_marker_dis_Euler_split} and \eqref{Eq_marker_dis_PC}, we find that in the
split method, the correction step is applied only to one velocity component, and the
correction term is not an arithmetic average of the velocity at two different locations.
A balanced split scheme is recovered by alternating in time the first direction of 1D 
advection. Still, the split scheme yields much smaller errors than the unsplit-Euler
method for linear and stationary velocity field components.

\subsection{Zalesak's disk}

The Zalesak's disk \cite{Zalesak_1979_31} test case is used to assess the
ability of the EBIT method to deal with sharp edges and corners.
A notched circular interface of radius $R = 0.15$ and center at $(0.5, 0.75)$
is placed inside the unit square domain. The notched width is $0.05$ and
its length is $0.25$. The velocity field and the timestep are both constant
and equal to those of the rotation test. 
The accuracy is measured by the area and symmetric difference errors.

% ----------
\begin{figure}
\begin{center}
\begin{tabular}{cc}
\includegraphics[width=0.45\textwidth]{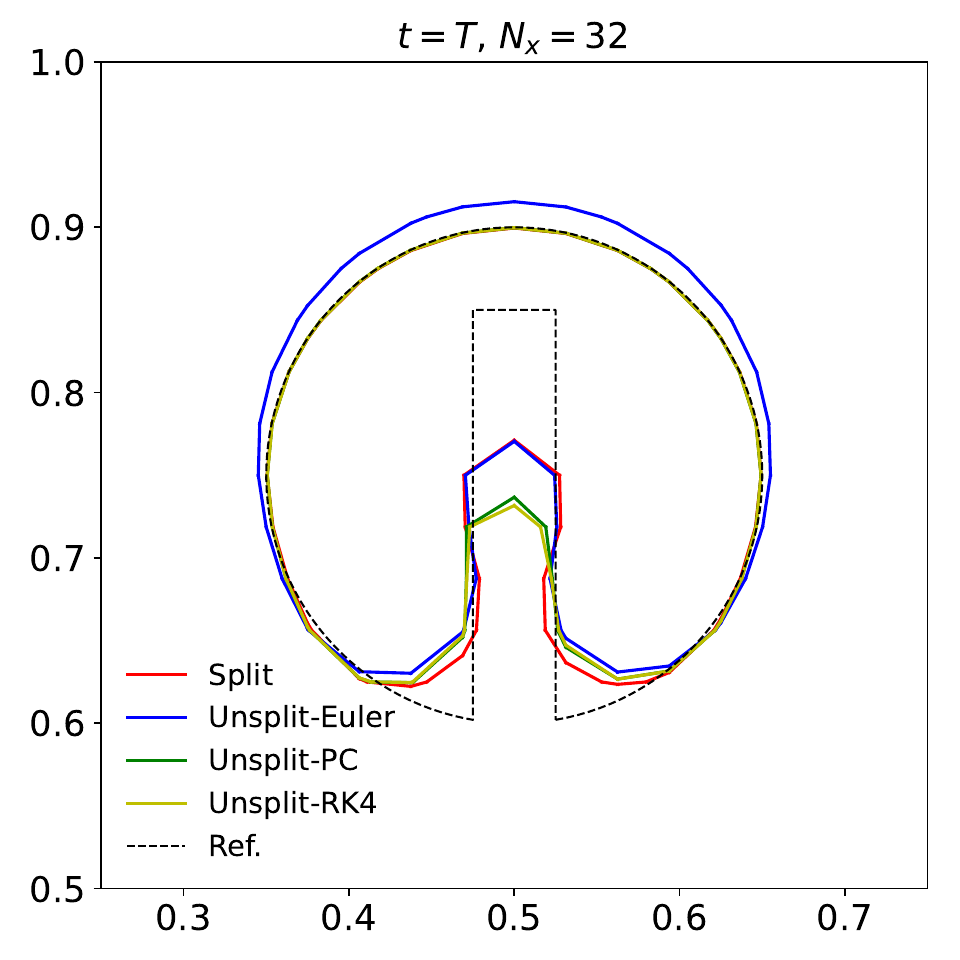} &
\includegraphics[width=0.45\textwidth]{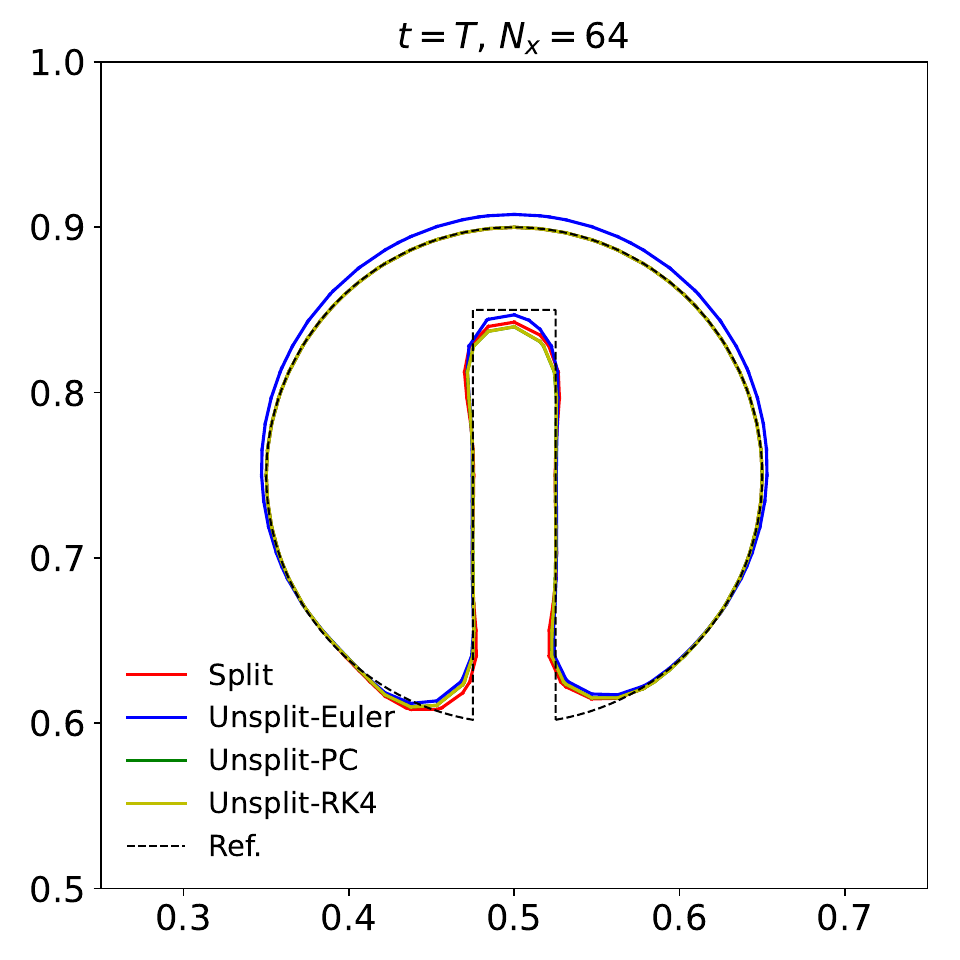}\\
(a) & (b) \\
\includegraphics[width=0.45\textwidth]{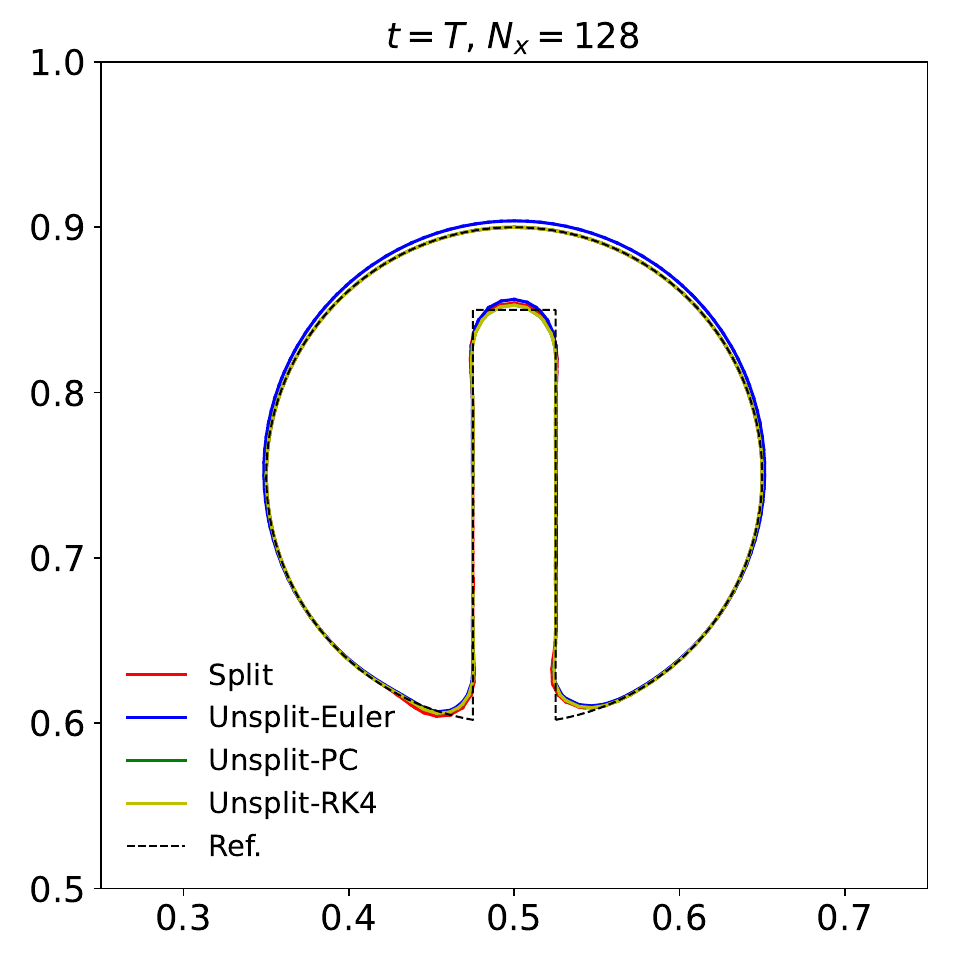} &
\includegraphics[width=0.45\textwidth]{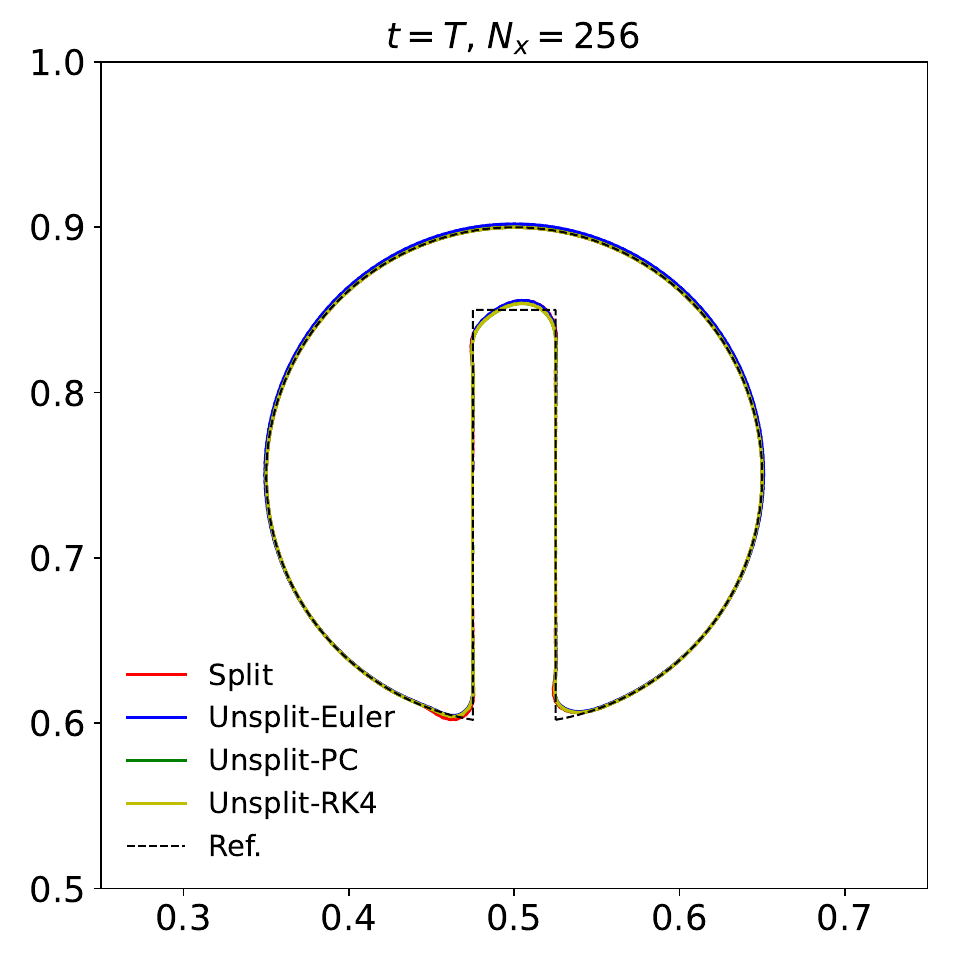}\\
(c) & (d) 
\end{tabular}
\end{center}
\caption{Interface lines at the end of Zalesak's disk test for different
grid resolutions: (a) $N_x = 32$; (b) $N_x = 64$; (c) $N_x = 128$; 
(d) $N_x = 256$}
\label{Fig_zalesak_intf}
\end{figure}
% ----------

% ----------
\begin{figure}
\begin{center}
\begin{tabular}{cc}
\includegraphics[width=0.45\textwidth]{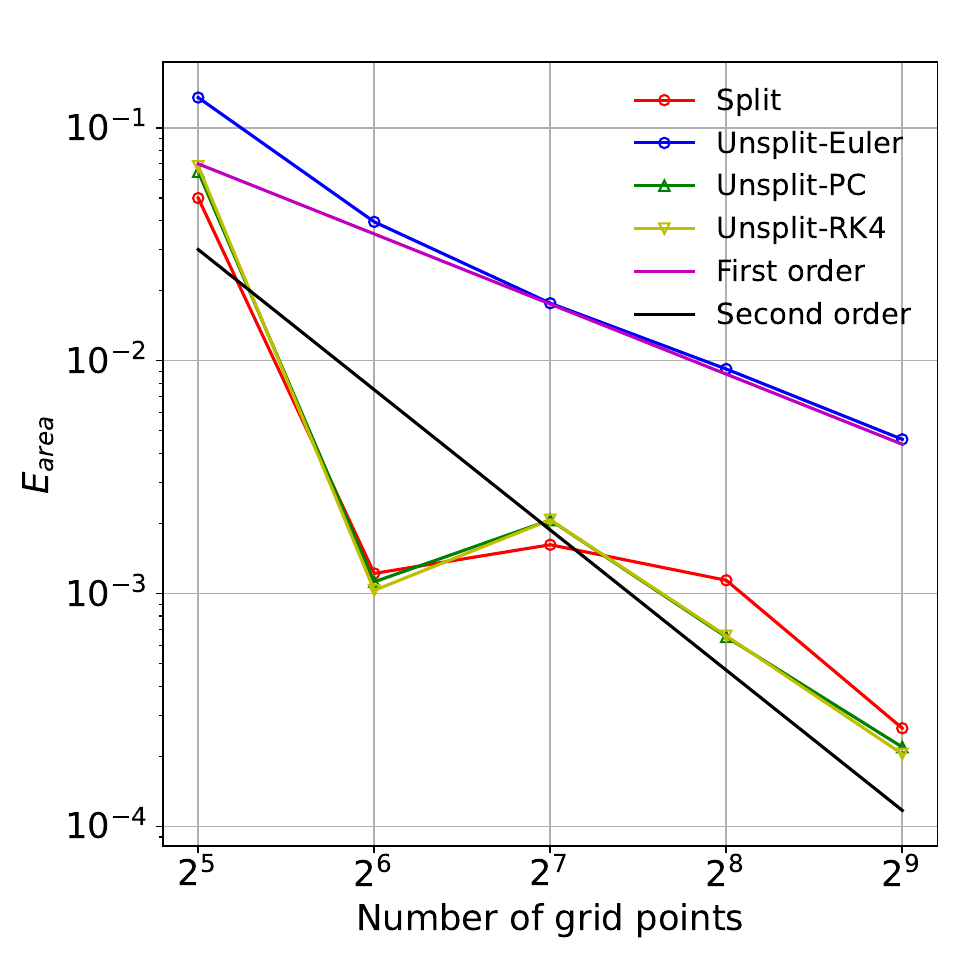} &
\includegraphics[width=0.45\textwidth]{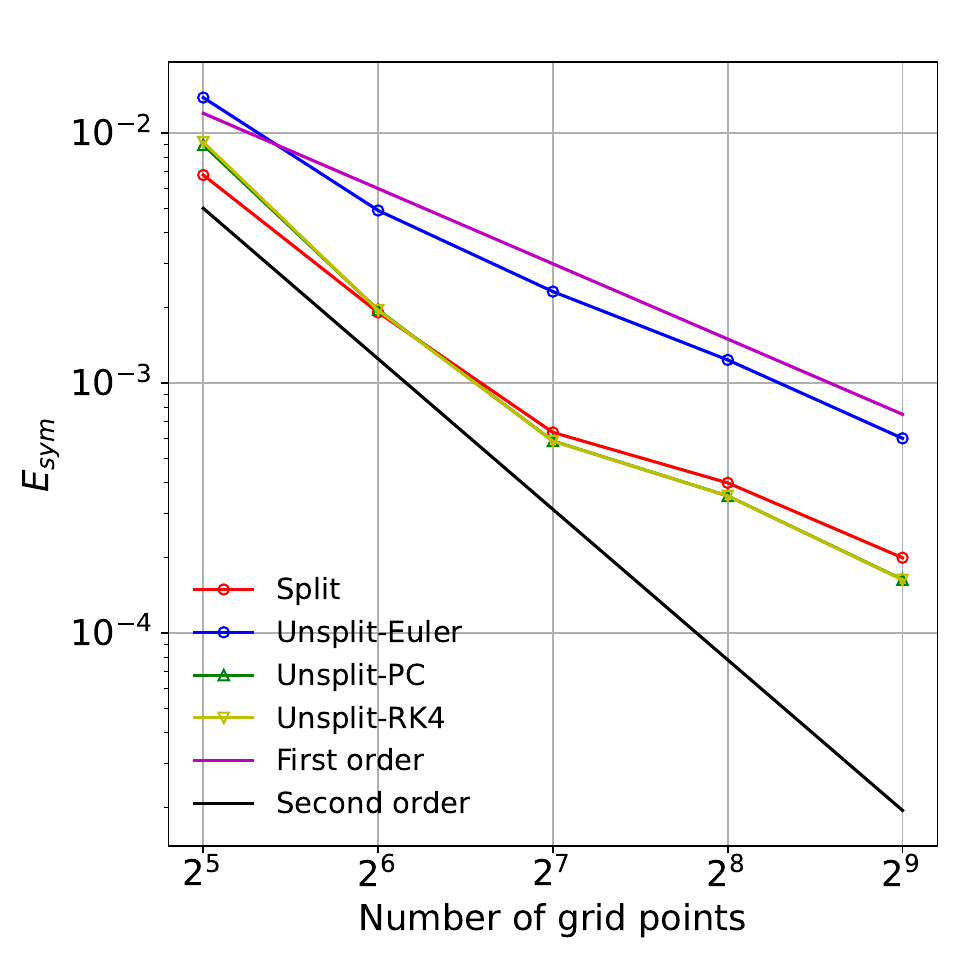}\\
(a) & (b)
\end{tabular}
\end{center}
\caption{Errors in Zalesak's disk test as a function of grid resolution:
(a) area error $E_{area}$; (b) symmetric difference error $E_{sym}$}
\label{Fig_zalesak_error}
\end{figure}
% ----------

% ----------
\begin{table}[hbt!]
\footnotesize
\caption{Mesh convergence study for Zalesak's disk test}
\centering
\begin{tabular}{cc|ccccc}
\hline 
 &$N_x$& 32 & 64 & 128 & 256 & 512\\ 
\hline 
Split &$E_{area}$ & $4.99 \times 10^{-2}$ & $1.22 \times 10^{-3}$ & $1.62 \times 10^{-3}$ & $1.14 \times 10^{-3}$ & $2.64 \times 10^{-4}$\\ 
&$E_{sym}$ & $6.80 \times 10^{-3}$ & $1.92 \times 10^{-3}$ & $6.35 \times 10^{-4}$ & $3.99 \times 10^{-4}$ & $2.00 \times 10^{-4}$\\
\hline 
Unsplit-Euler & $E_{area}$ & $1.35 \times 10^{-1}$ & $3.94 \times 10^{-2}$ & $1.76 \times 10^{-2}$ & $9.20 \times 10^{-3}$ & $4.59 \times 10^{-3}$\\
&$E_{sym}$ & $1.39 \times 10^{-2}$ & $4.90 \times 10^{-3}$ & $2.32 \times 10^{-3}$ & $1.24 \times 10^{-3}$ & $6.01 \times 10^{-4}$\\
\hline 
Unsplit-PC &$E_{area}$ & $6.48 \times 10^{-2}$ & $1.12 \times 10^{-3}$ & $2.07 \times 10^{-3}$ & $ 6.51 \times 10^{-4}$ & $ 2.19 \times 10^{-4}$\\ 
&$E_{sym}$ & $8.97 \times 10^{-3}$ & $1.96 \times 10^{-3}$ & $5.87 \times 10^{-4}$ & $3.54 \times 10^{-4}$ & $1.64 \times 10^{-4}$\\
\hline 
Unsplit-RK4 &$E_{area}$ & $6.82 \times 10^{-2}$ & $1.03 \times 10^{-3}$ & $2.07 \times 10^{-3}$ & $6.58 \times 10^{-4}$ & $ 2.05 \times 10^{-4}$\\ 
&$E_{sym}$ & $9.20 \times 10^{-3}$ & $1.95 \times 10^{-3}$ & $5.86 \times 10^{-4}$ & $3.54 \times 10^{-4}$ & $1.63 \times 10^{-4}$\\
\hline 
\end{tabular}
\label{Tab_zalesak_error}
\normalsize
\end{table}
% ----------

The interface lines obtained with different advection methods are presented
in Fig.~\ref{Fig_zalesak_intf}, at the end of the simulation and for
different mesh resolutions. 
Away from the notch, the results are similar to those of the rotation test,
as expected. In the areas near the four corners, the unsplit methods typically 
show a higher degree of smoothing at coarser mesh resolutions. However, as the mesh
resolution increases, the level of smoothing becomes comparable across all the 
methods being evaluated.
The results obtained with the unsplit-PC and unsplit-RK4 methods do not
show appreciable differences between them.

The area error $E_{area}$ and the symmetric difference error $E_{sym}$ are
listed in Table~\ref{Tab_zalesak_error} and are shown in
Fig.~\ref{Fig_zalesak_error}. 
For the unsplit-Euler method, a first-order convergence rate is still
observed for both errors. 
For the split method and the unsplit-PC and unsplit-RK4 methods, the two
errors are quite similar. The area error oscillates at the lowest
resolutions, and then approaches second-order convergence as the mesh
resolution increases. 
Due to the presence of sharp corners, the order of convergence for the symmetric
error of these three methods initially approaches second order but
then shifts to nearly first order. 
Additionally, the two high-order unsplit methods do not provide greater accuracy
than the split method, which is a contrasting outcome compared to the rotation test.
In other words, in Zalesak’s disk test, the smoothing of the interface near the
corners contributes most significantly to the symmetric error.

\subsection{Single vortex}

The single vortex test was designed to test the ability of an interface
tracking method to follow the evolution in time of an interface that is
first highly stretched and deformed \cite{Rider_1998_141}. 
A circular interface of radius $R=0.15$ and
center at $(0.5, 0.75)$ is placed inside the unit square domain. 
A divergence-free velocity field $(u, v) = (\partial \phi \big/ \partial y,
- \partial \phi \big/ \partial x)$, described by the stream function 
$\phi = \pi^{-1} \sin^2(\pi x) \sin^2(\pi y) \cos(\pi t/T)$, 
is imposed. The cosinusoidal time dependence slows down the flow, with
the maximum deformation occurring at $t = 0.5\,T$, then reverses the flow
and the interface returns to its initial position, without distortion at
$t = T$. Furthermore, as the value of the period $T$ increases, a thinner
and thinner revolving ligament develops.

It should be noted that, in addition to the time integration and the
interface reconstruction steps, the bilinear interpolation, which is used
to calculate the velocity at the marker position, could also introduce
some approximation error in this test because the 
velocity field components are nonlinear functions of the position.

The timestep is kept constant in a simulation and is computed from the
maximum horizontal component of the velocity at time $t=0$, $u_{max}$, 
so that $\textrm{CFL} = u_{max} \,\Delta t / h = 0.125$. The accuracy
of the advection methods is again measured by the area, shape, and symmetric
difference errors.

To obtain the reference solution for comparison, we place an ordered list of $512$ Lagrangian
markers on the initial circular interface, connect them with segments, and advect
them numerically along the flow streamlines. A RK4 method, with an adaptive
timestep in the SciPy Python library, is used to solve the system of two ordinary
differential equations $d \X \big/ d t = \U(x(t), y(t), t)$, which describes the motion of the
markers. A user-defined maximum timestep, $\Delta t_{max} = 0.01$, is used
in the numerical integration.

% ----------
\begin{figure}
\begin{center}
\begin{tabular}{cc}
\includegraphics[width=0.45\textwidth]{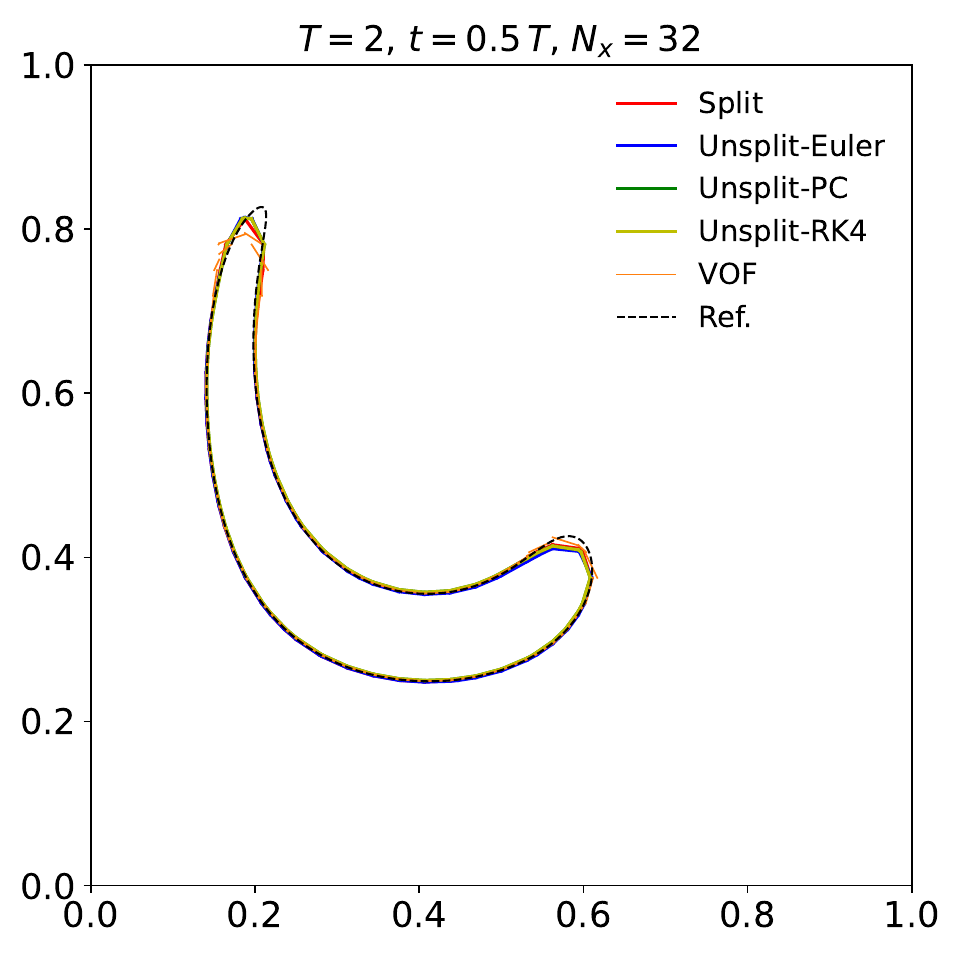} &
\includegraphics[width=0.45\textwidth]{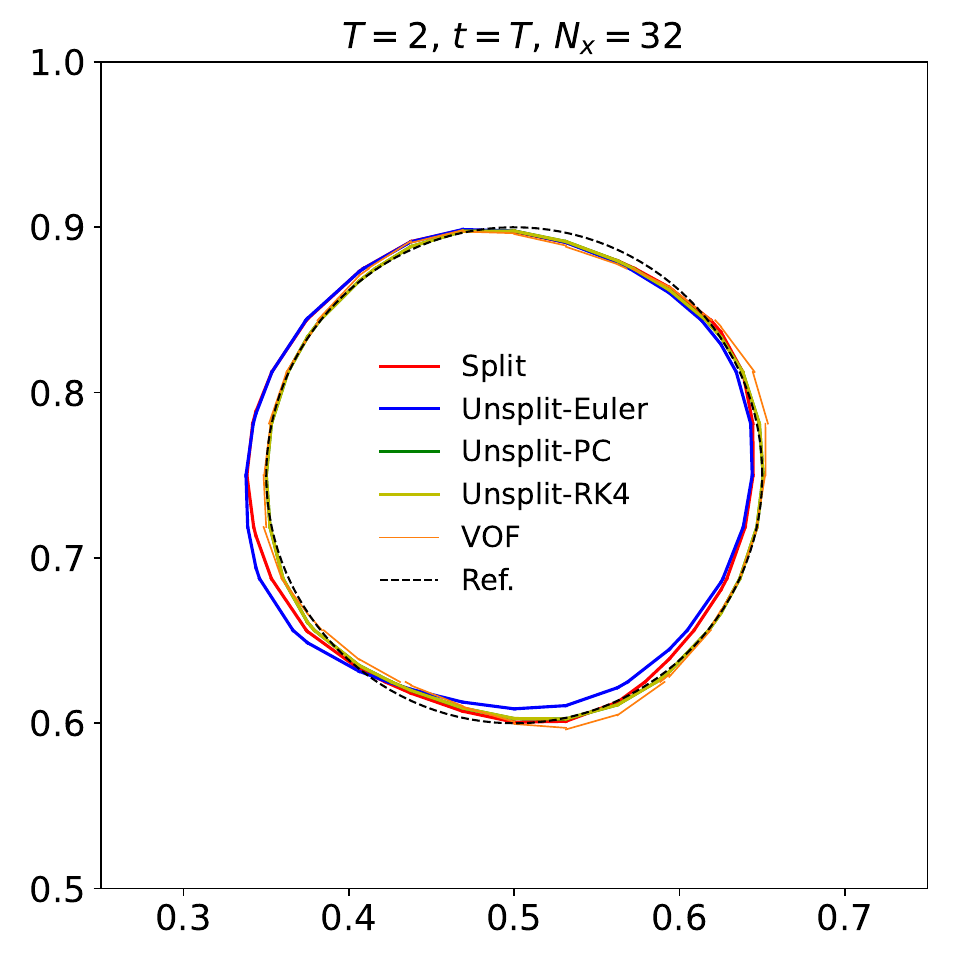}\\
(a) & (b) \\
\includegraphics[width=0.45\textwidth]{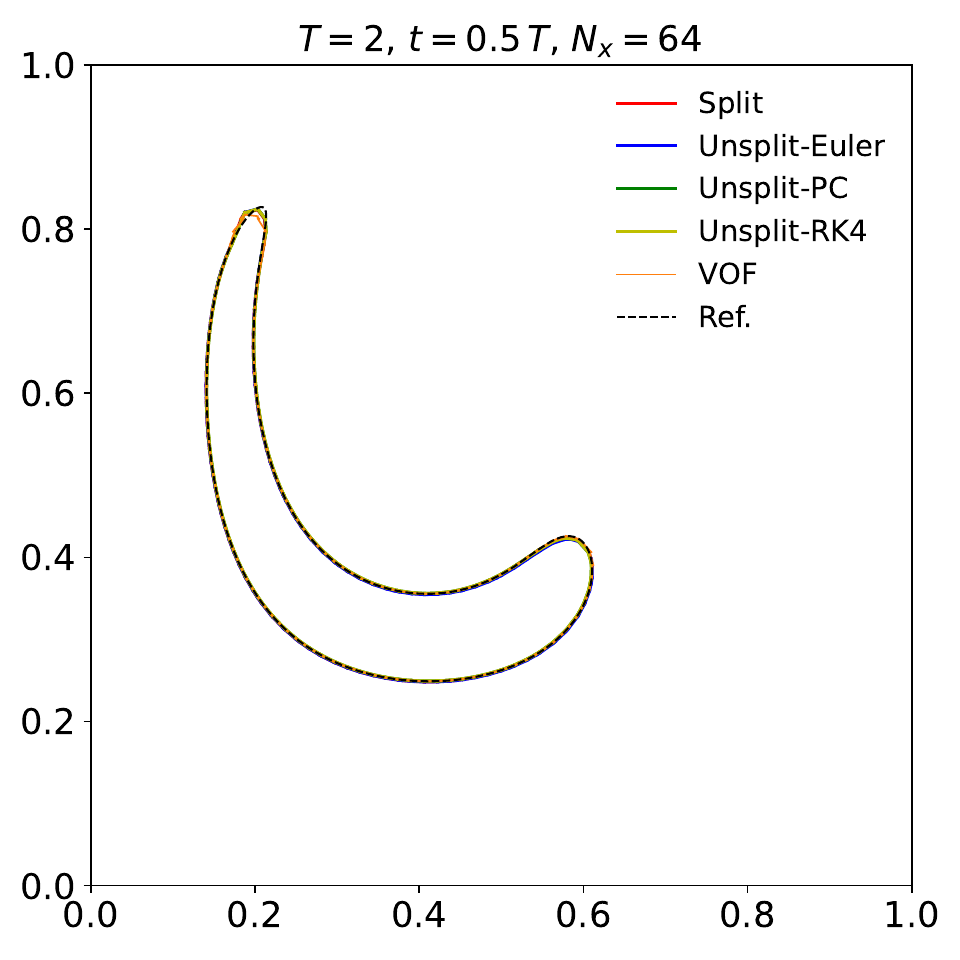} &
\includegraphics[width=0.45\textwidth]{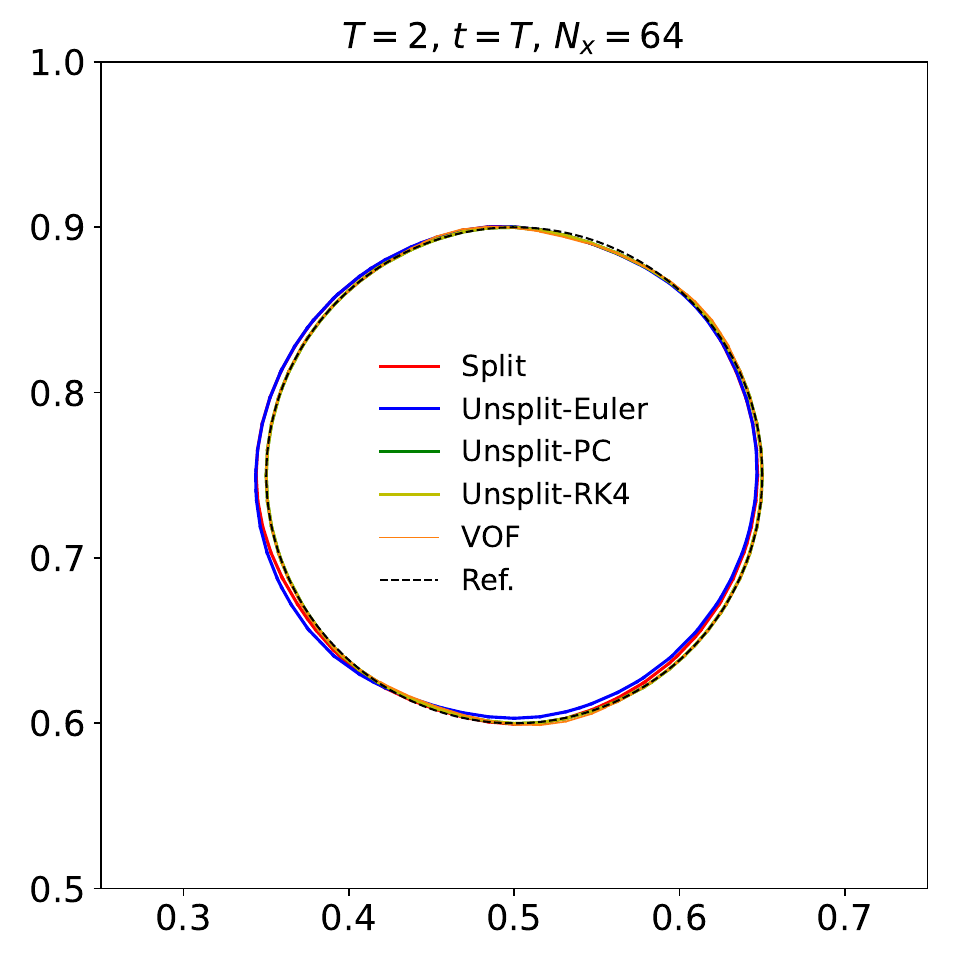}\\
(c) & (d) 
\end{tabular}
\end{center}
\caption{Interface lines at half time and the end of the single vortex 
test with period $T = 2$, for different grid resolutions: (a) $t = 0.5\,T$,
$N_x = 32$; (b) $t = T$, $N_x = 32$; (c) $t = 0.5\,T$, $N_x = 64$; 
(d) $t = T$, $N_x = 64$}
\label{Fig_vortex_intf_t2_1}
\end{figure}

% ----------
\begin{figure}
\begin{center}
\begin{tabular}{cc}
\includegraphics[width=0.45\textwidth]{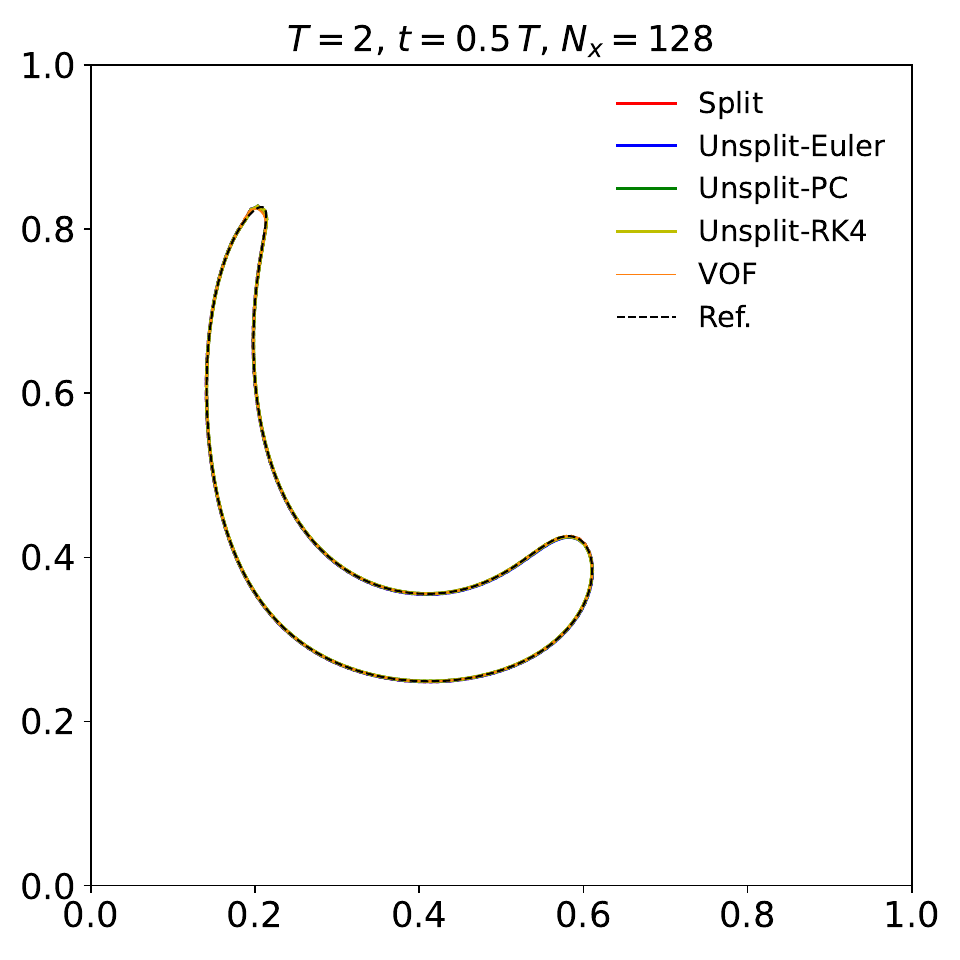} &
\includegraphics[width=0.45\textwidth]{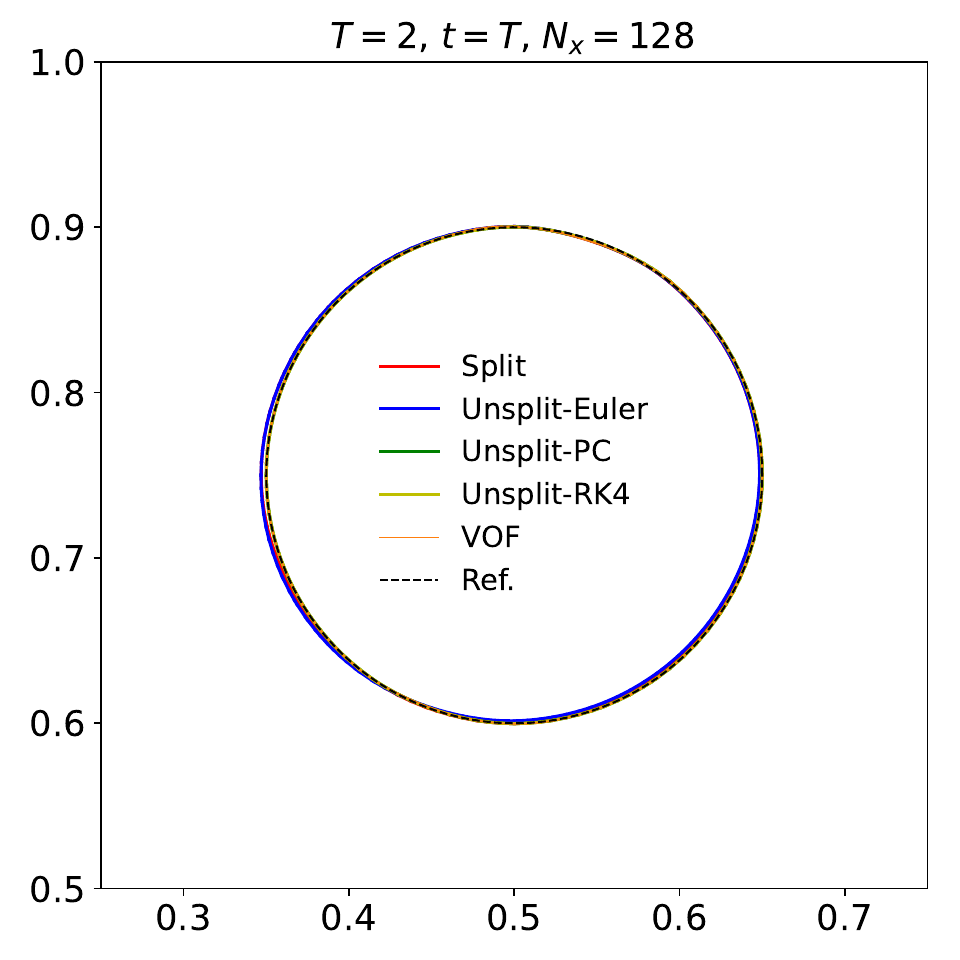}\\
(a) & (b) \\
\includegraphics[width=0.45\textwidth]{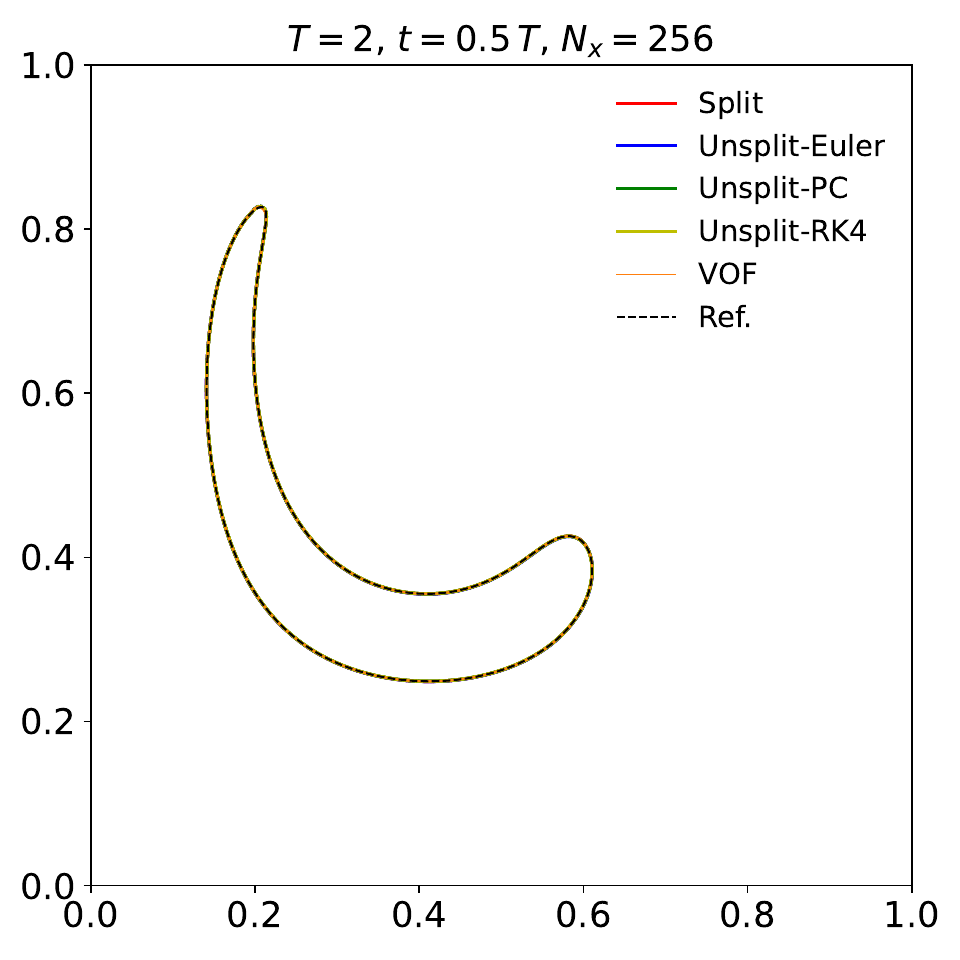} &
\includegraphics[width=0.45\textwidth]{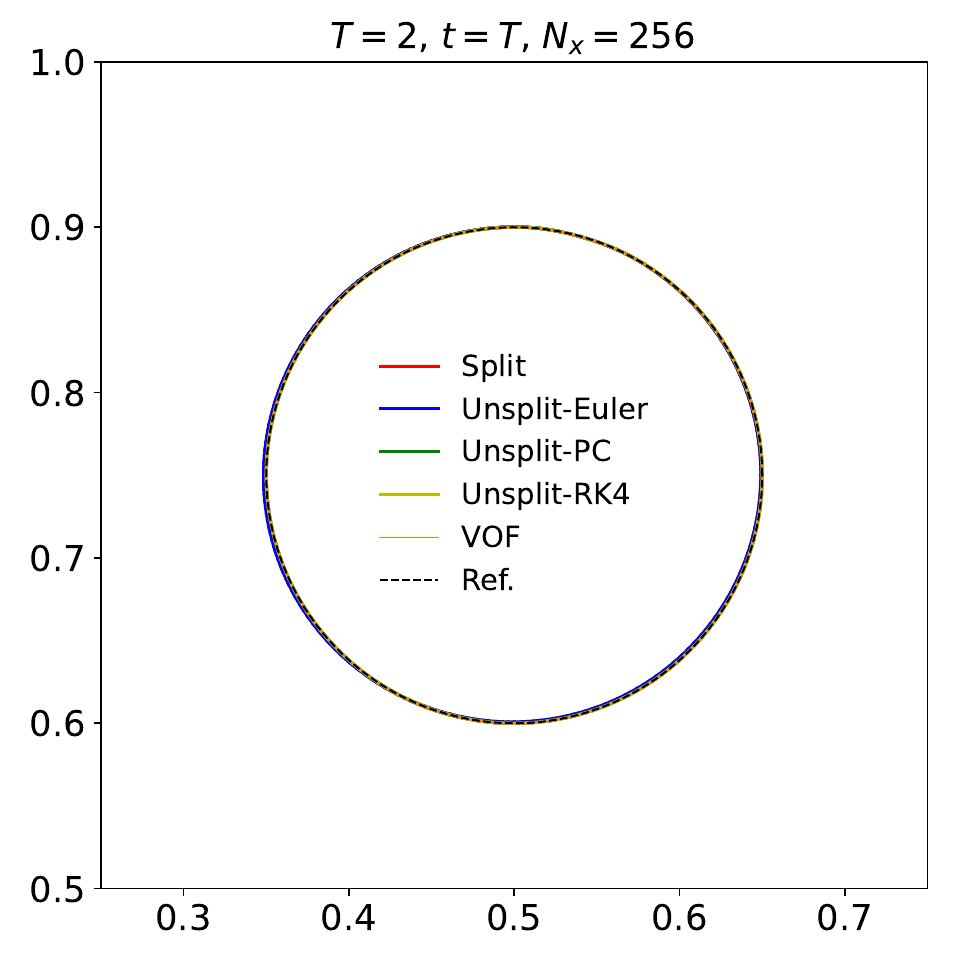}\\
(c) & (d) 
\end{tabular}
\end{center}
\caption{Interface lines at half time and the end of the single vortex 
test with period $T = 2$, for different grid resolutions: (a) $t = 0.5\,T$,
$N_x = 128$; (b) $t = T$, $N_x = 128$; (c) $t = 0.5\,T$, $N_x = 256$; 
(d) $t = T$, $N_x = 256$}
\label{Fig_vortex_intf_t2_2}
\end{figure}
% ----------

\begin{table}[hbt!]
\footnotesize
\caption{Mesh convergence study for the single vortex test with period 
$T = 2$}
\centering
\begin{tabular}{cc|ccccc}
\hline 
 &$N_x$& 32 & 64 & 128 & 256 & 512 \\ 
\hline 
Split & $E_{area}$ & $1.15\times 10^{-2}$ & $5.37 \times 10^{-3}$ & $2.62 \times 10^{-3}$ & $1.25 \times 10^{-3}$ & $6.10 \times 10^{-4}$ \\ 
& $E_{shape}$ & $1.17 \times 10^{-2}$ & $6.13 \times 10^{-3}$ & $3.07 \times 10^{-3}$ & $1.54 \times 10^{-3}$ & $7.69 \times 10^{-4}$ \\
& $E_{sym}$ & $4.59 \times 10^{-3}$ & $2.16 \times 10^{-3}$ & $1.06 \times 10^{-3}$ & $5.27 \times 10^{-4}$ & $2.62 \times 10^{-4}$ \\
\hline
Unsplit-Euler & $E_{area}$ & $5.34 \times 10^{-3}$ & $4.96 \times 10^{-4}$ & $9.30 \times 10^{-4}$ & $4.74 \times 10^{-4}$ & $2.30 \times 10^{-4}$\\
& $E_{shape}$ & $1.62 \times 10^{-2}$ & $6.67 \times 10^{-3}$ & $3.35 \times 10^{-3}$ & $1.67 \times 10^{-3}$ & $8.36 \times 10^{-4}$ \\
& $E_{sym}$ & $6.76 \times 10^{-3}$ & $3.10 \times 10^{-3}$ & $1.50 \times 10^{-3}$ & $7.45 \times 10^{-4}$ & $3.71 \times 10^{-4}$ \\
\hline 
Unsplit-PC & $E_{area}$ & $\underline{7.71\times 10^{-3}}$ & $1.01 \times 10^{-3}$ & $\underline{4.12 \times 10^{-5}}$ & $4.81 \times 10^{-5}$ & $\underline{1.65 \times 10^{-5}}$\\ 
& $E_{shape}$ & $6.04 \times 10^{-3}$ & $2.11 \times 10^{-3}$ & $5.99 \times 10^{-4}$ & $1.88 \times 10^{-4}$ & $5.52 \times 10^{-5}$\\
& $E_{sym}$ & $1.72 \times 10^{-3}$ & $\underline{3.36 \times 10^{-4}}$ & $\underline{6.72 \times 10^{-5}}$ & $1.51 \times 10^{-5}$ & $3.02 \times 10^{-6}$\\
\hline
Unsplit-RK4 & $E_{area}$ & $\underline{7.08\times 10^{-3}}$ & $1.01 \times 10^{-3}$ & $\underline{4.38 \times 10^{-5}}$ & $4.81 \times 10^{-5}$ & $\underline{1.64 \times 10^{-5}}$\\ 
& $E_{shape}$ & $6.04 \times 10^{-3}$ & $2.11 \times 10^{-3}$ & $5.99 \times 10^{-4}$ & $1.88 \times 10^{-4}$ & $5.52 \times 10^{-5}$\\
& $E_{sym}$ & $1.72 \times 10^{-3}$ & $\underline{3.34 \times 10^{-4}}$ & $\underline{6.73 \times 10^{-5}}$ & $1.51 \times 10^{-5}$ & $3.02 \times 10^{-6}$\\
\hline
VOF & $E_{shape}$ & $8.79 \times 10^{-3}$ & $3.00 \times 10^{-3}$ & $1.17 \times 10^{-3}$ & $4.11 \times 10^{-4}$ & $1.21 \times 10^{-4}$\\
& $E_{sym}$ & $3.16 \times 10^{-3}$ & $6.93 \times 10^{-4}$ & $1.43 \times 10^{-4}$ & $3.14 \times 10^{-5}$ & $7.48 \times 10^{-6}$\\
\hline 
\end{tabular}
\label{Tab_vortex_error_t2}
\normalsize
\end{table}
% ----------
\begin{figure}
\begin{center}
\begin{tabular}{ccc}
\includegraphics[width=0.33\textwidth]{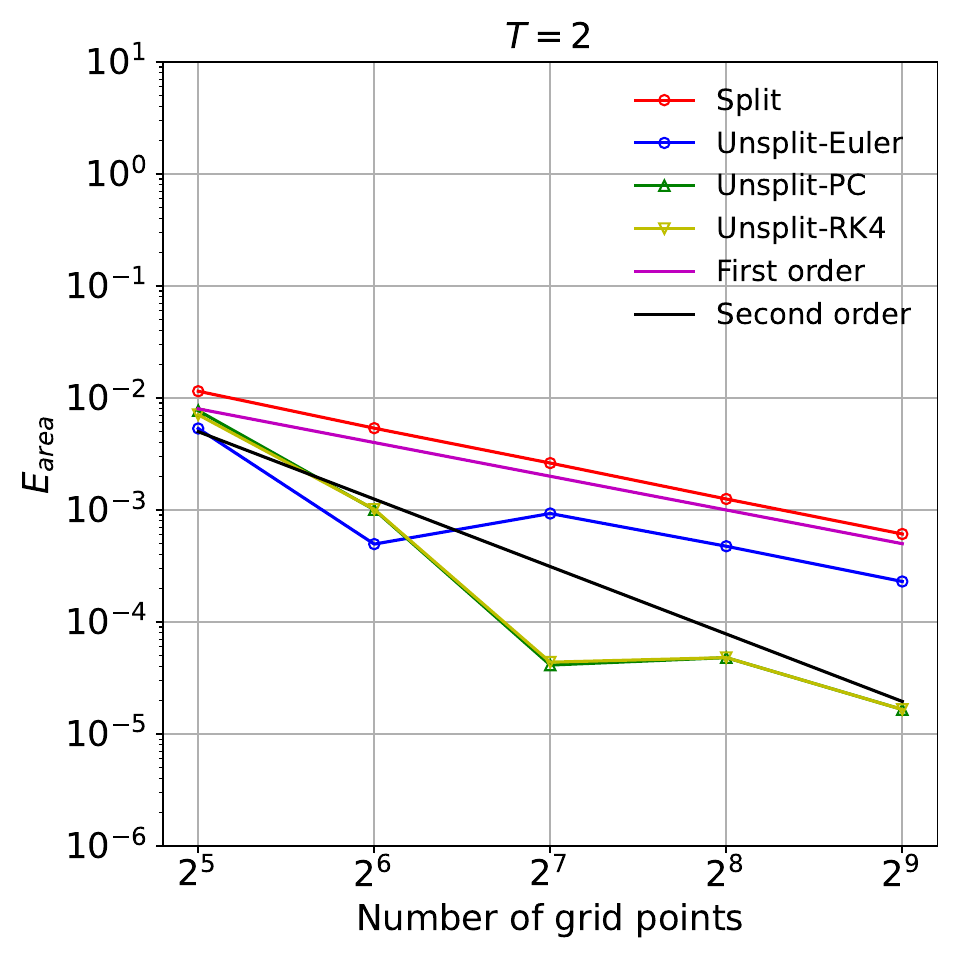} &
\includegraphics[width=0.33\textwidth]{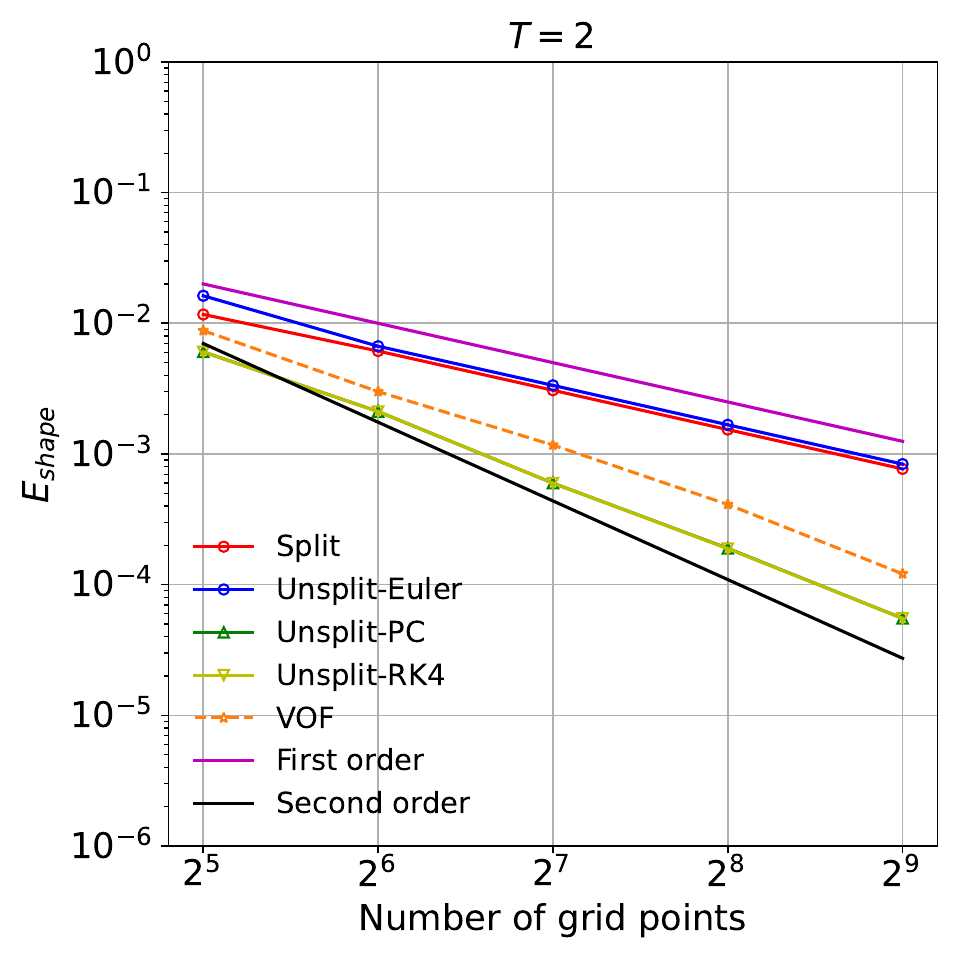} &
\includegraphics[width=0.33\textwidth]{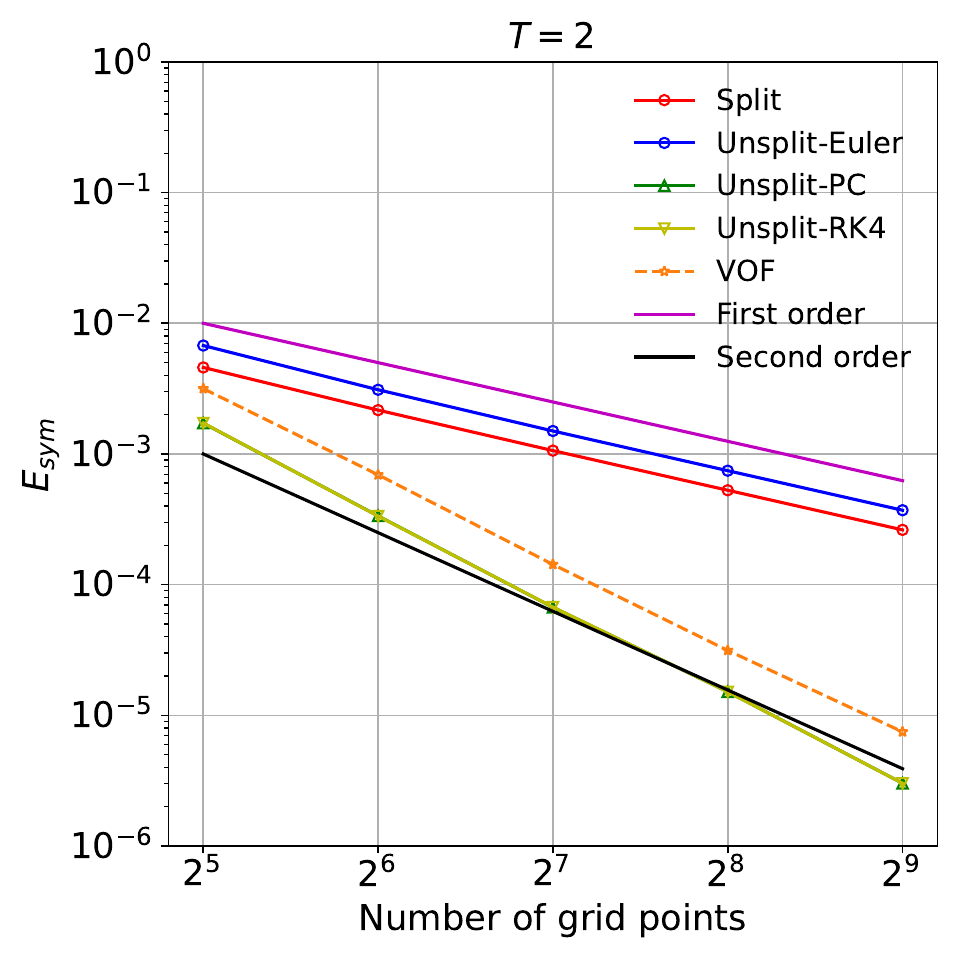}\\
(a) & (b) & (c)
\end{tabular}
\end{center}
\caption{Errors in the single vortex test with period $T=2$ as a function 
of grid resolution: (a) area error $E_{area}$; (b) shape error $E_{shape}$;
(c) symmetric difference error $E_{sym}$}
\label{Fig_vortex_error_t2}
\end{figure}
% ----------

The interface lines are shown in Figs.~\ref{Fig_vortex_intf_t2_1} and
\ref{Fig_vortex_intf_t2_2}, for different mesh resolutions, at their maximum
deformation and once they are back in their starting position. 
The results obtained with the split method and unsplit-Euler method present
a similar deviation from the reference solution. The unsplit-PC and
unsplit-RK4 methods better recover the circular shape of the interface line
at the end of the simulation for all mesh resolutions.

The area error $E_{area}$, the shape error $E_{shape}$ and the symmetric
difference error $E_{sym}$ are listed in Table~\ref{Tab_vortex_error_t2} 
and are shown in Fig.~\ref{Fig_vortex_error_t2}, for the different methods
considered here. For the split and unsplit-Euler methods, first-order 
convergence with grid resolution is observed for all errors. 
The unsplit-Euler method presents smaller area errors, which may be caused
by fewer reconstruction steps, but larger shape and symmetric errors.

For the unsplit-PC and unsplit-RK4 methods, second-order convergence is
observed for all errors. We remark that the higher-order 
unsplit-RK4 method does not lead to smaller errors as in the rotation test.
The rare cases where the difference in errors is within three significant figures are
underlined in Table~\ref{Tab_vortex_error_t2}.
In this test, we can assess the relevance of bilinear interpolation in relation
to the measured errors. We have advected the markers using the analytical expression
of the velocity field without any interpolation, and we found that there is no
significant change in the results.
We can conclude that the dominant error is the reconstruction
error.

% ----------
\begin{figure}
\begin{center}
\begin{tabular}{cc}
\includegraphics[width=0.45\textwidth]{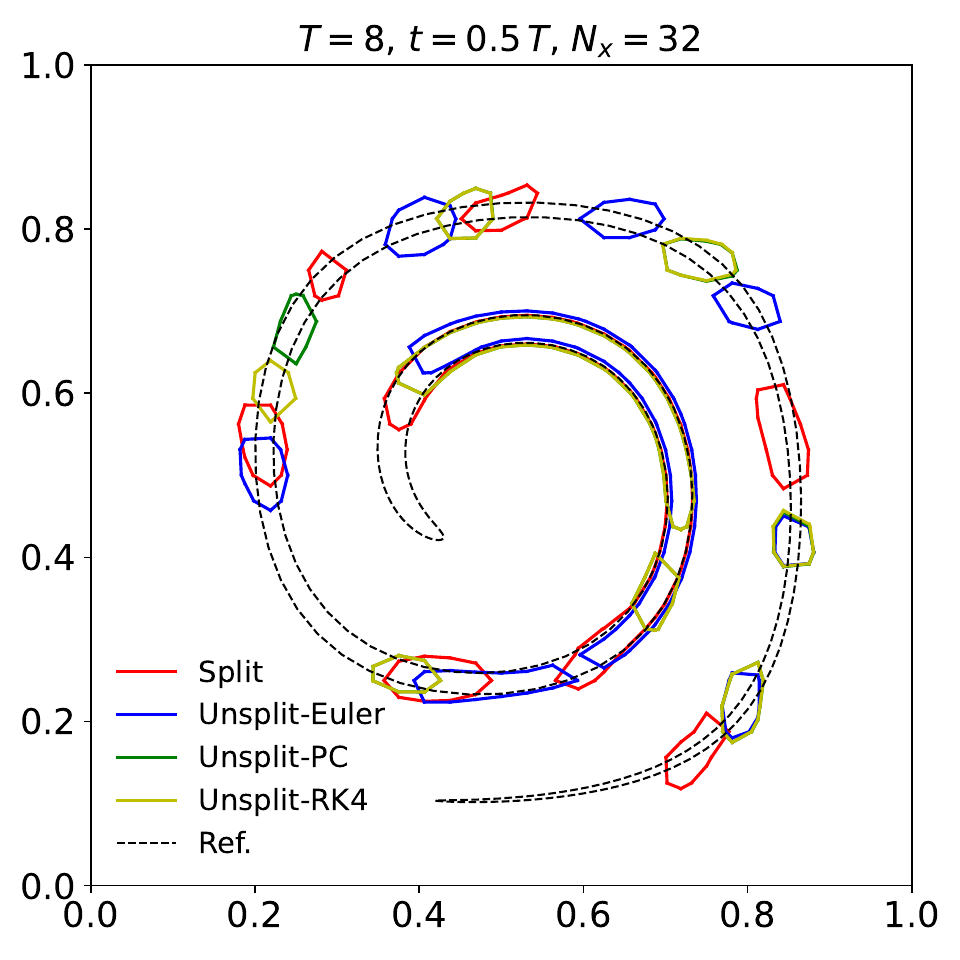} &
\includegraphics[width=0.45\textwidth]{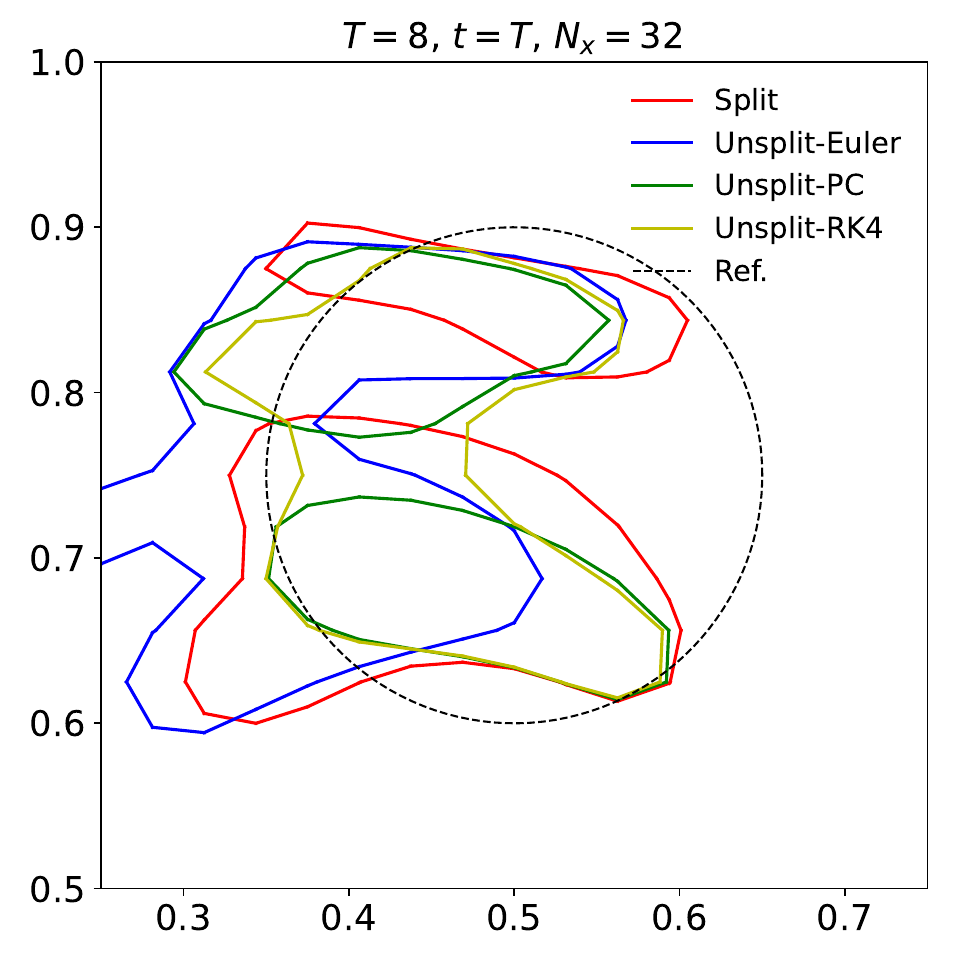}\\
(a) & (b)\\
\includegraphics[width=0.45\textwidth]{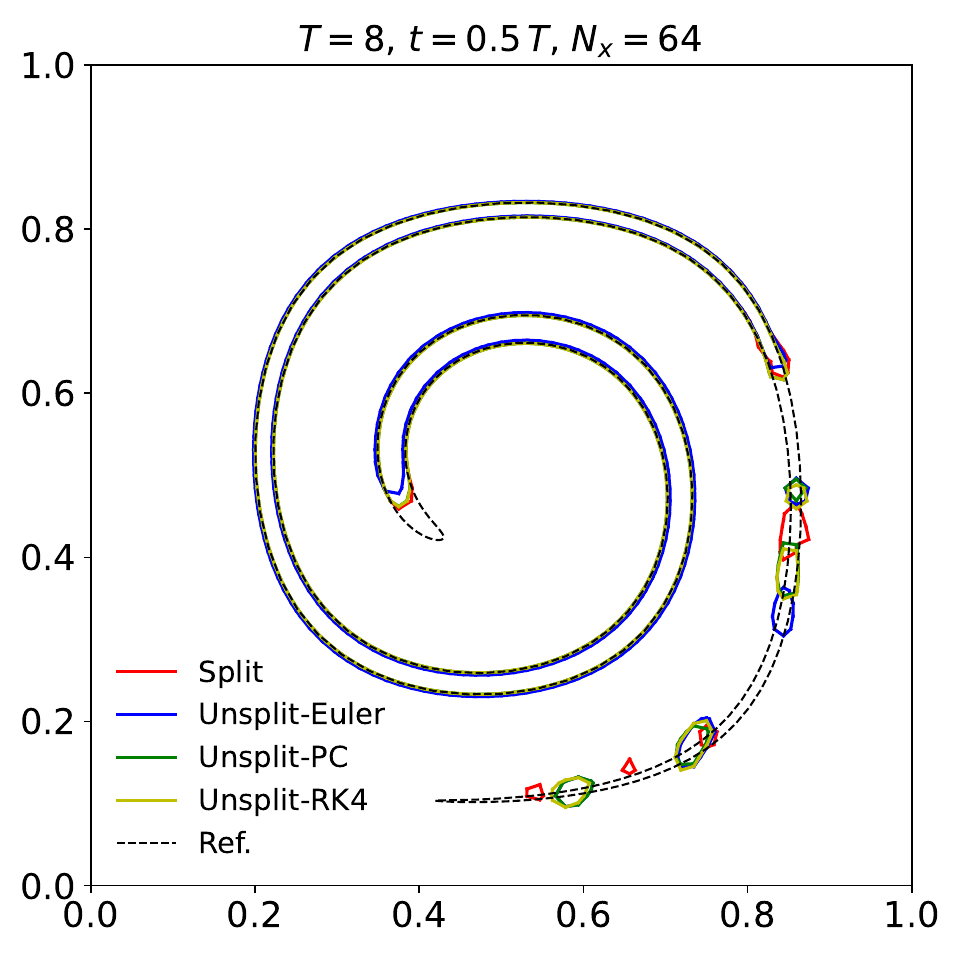} &
\includegraphics[width=0.45\textwidth]{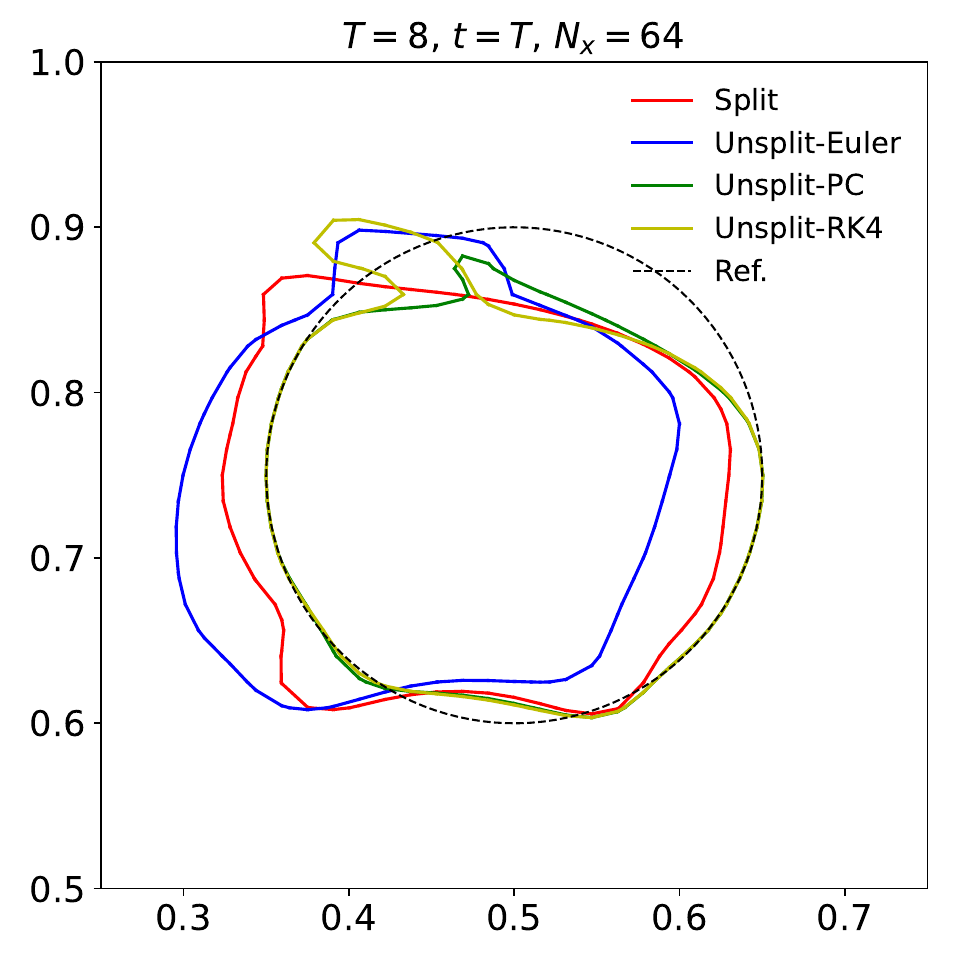}\\
(c) & (d)
\end{tabular}
\end{center}
\caption{Interface lines at half time and the end of the single vortex test with period
$T = 8$, for different grid resolutions: (a) $t = 0.5 T, N_x = 32$; (b) $t = T, N_x = 32$; (c) $t = 0.5 T, N_x = 64$; (d) $t = T, N_x = 64$}
\label{Fig_vortex_intf_t8_1}
\end{figure}
% ----------
\begin{figure}
\begin{center}
\begin{tabular}{cc}
\includegraphics[width=0.45\textwidth]{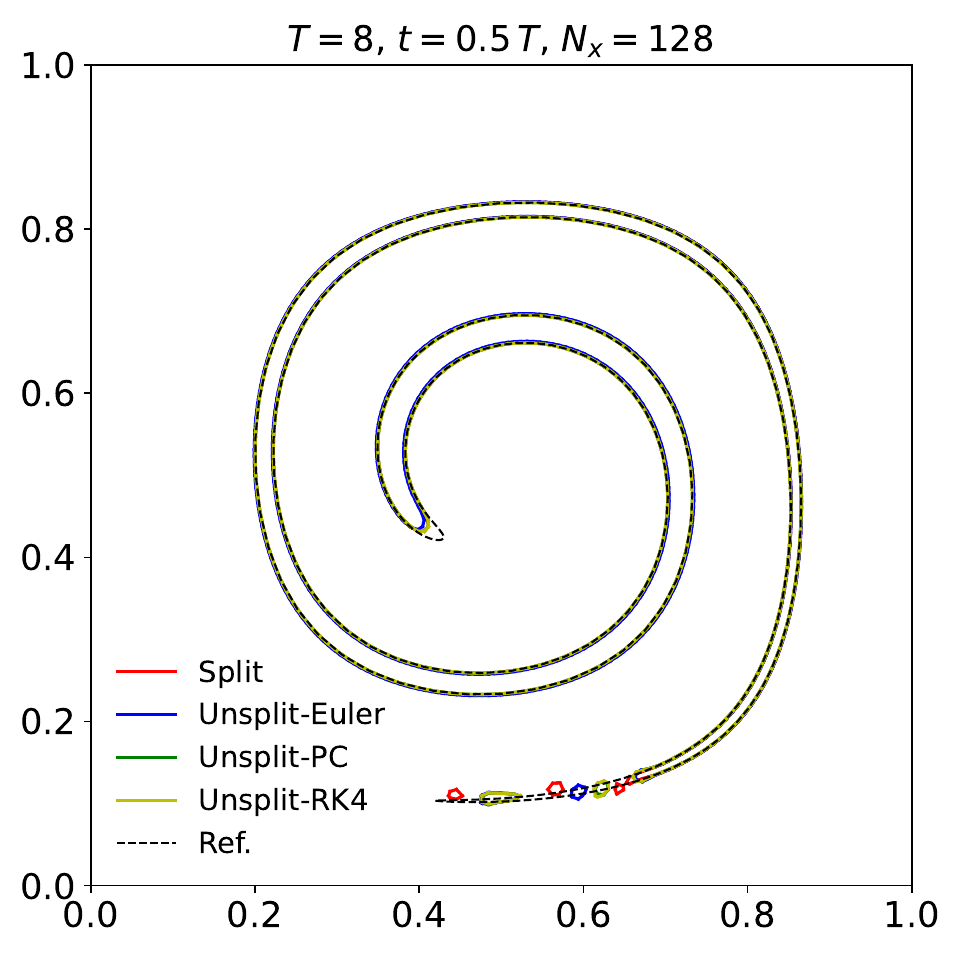} &
\includegraphics[width=0.45\textwidth]{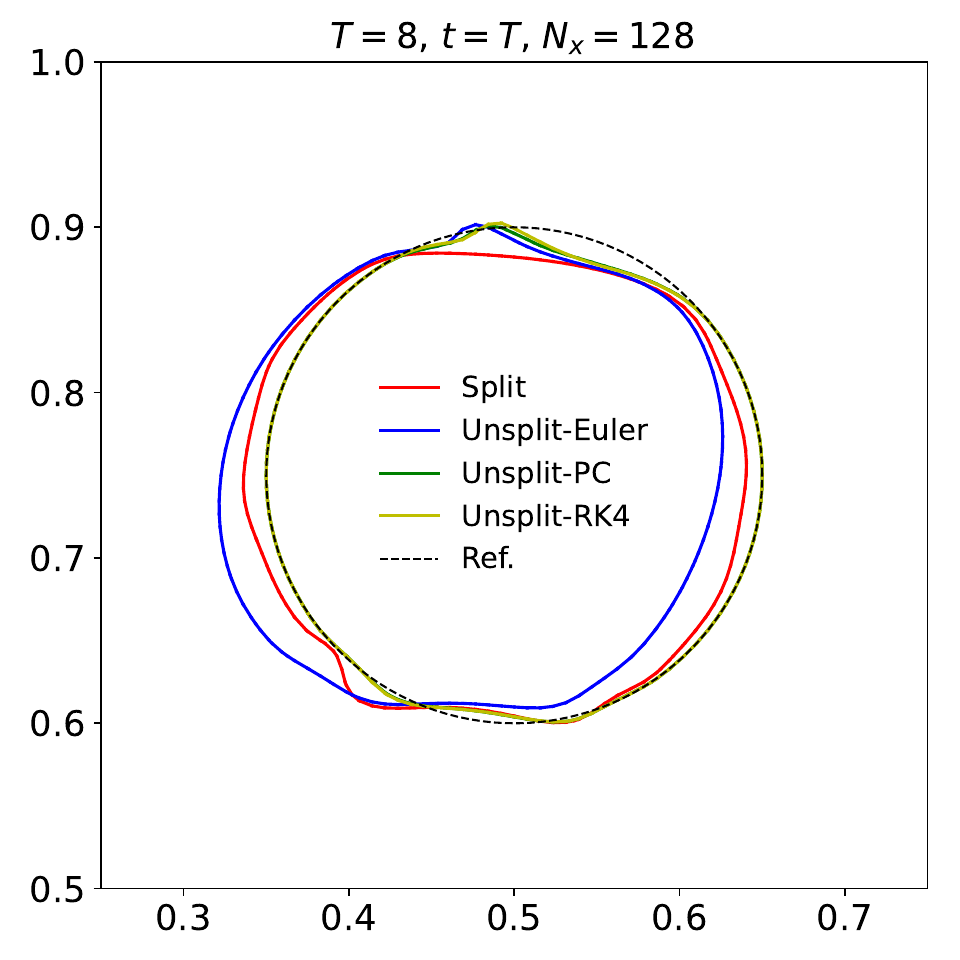}\\
(a) & (b)\\
\includegraphics[width=0.45\textwidth]{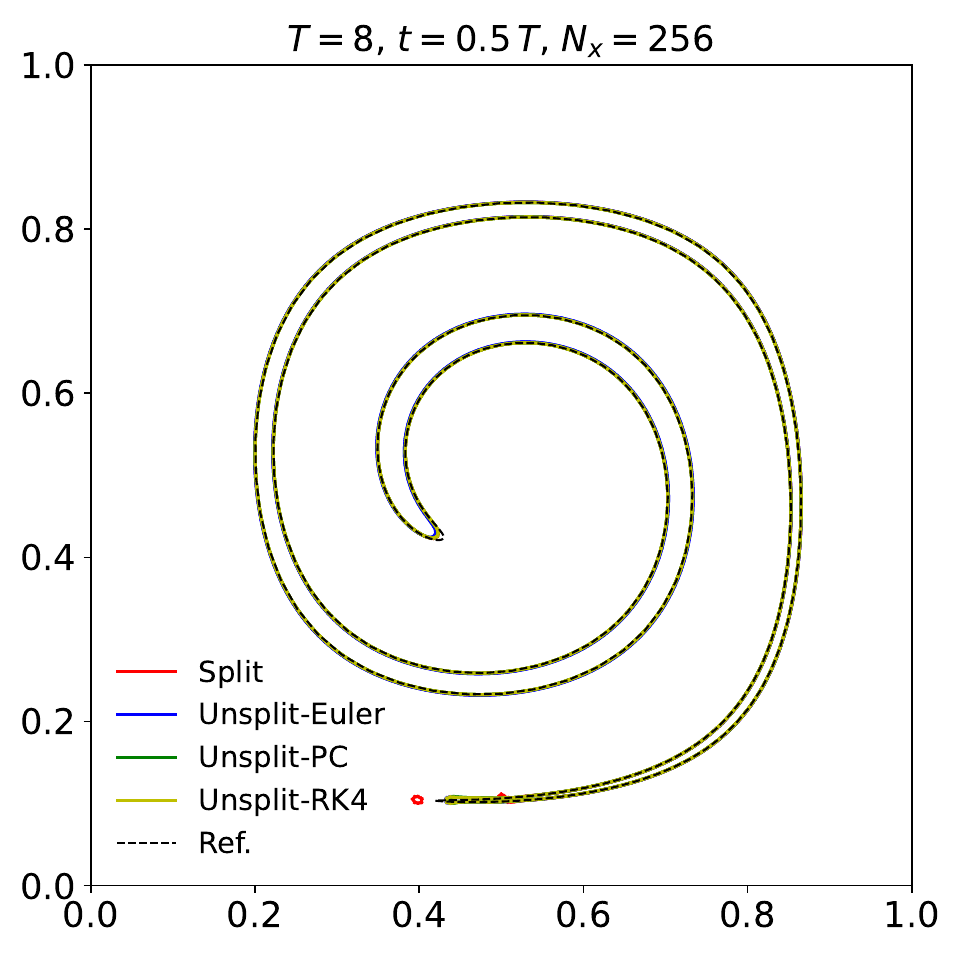} &
\includegraphics[width=0.45\textwidth]{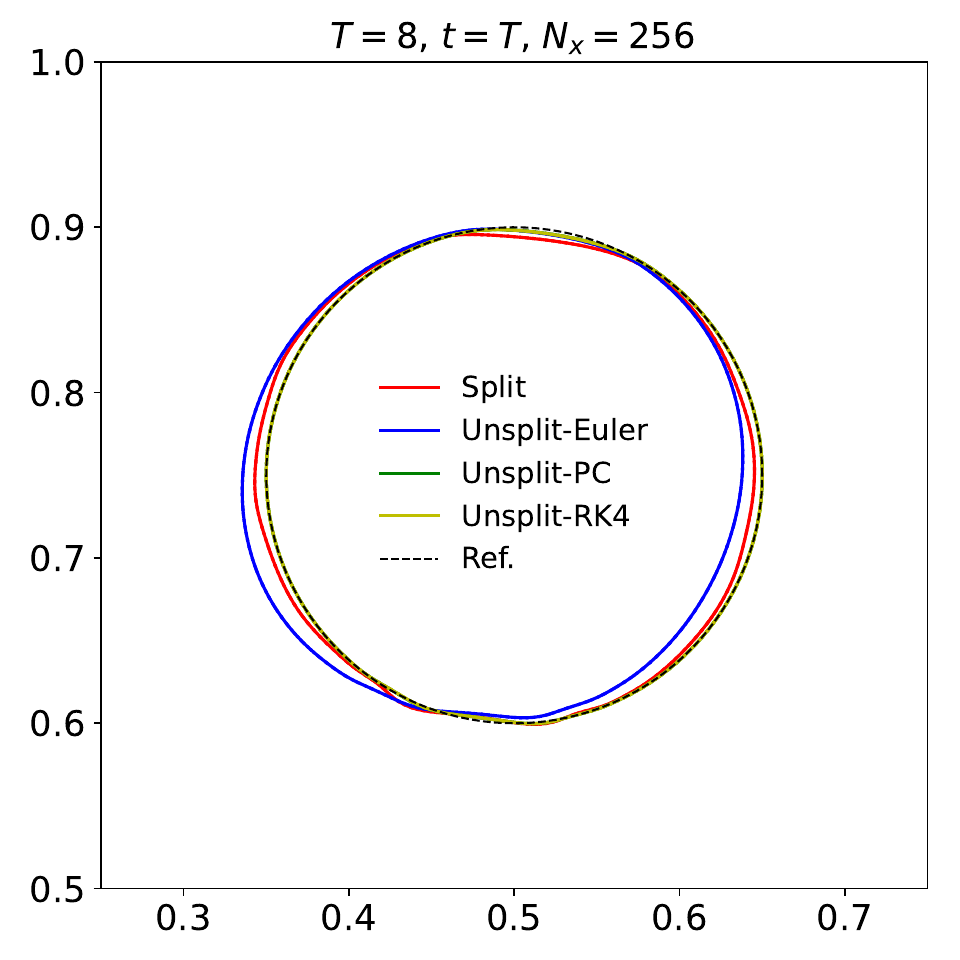}\\
(c) & (d)
\end{tabular}
\end{center}
\caption{Interface lines at half time and the end of the single vortex test with period
$T = 8$, for different grid resolutions: (a) $t = 0.5 T, N_x = 128$; (b) $t = T, N_x = 128$; (c) $t = 0.5 T, N_x = 256$; (d) $t = T, N_x = 256$}
\label{Fig_vortex_intf_t8_2}
\end{figure}
% ----------

\begin{table}[hbt!]
\footnotesize
\caption{Mesh convergence study for the single vortex test with period $T = 8$}
\centering
\begin{tabular}{cc|cccccc}
\hline 
 &$N_x$& 32 & 64 & 128 & 256 & 512 & 1024\\ 
\hline 
Split &$E_{area}$ & $2.87 \times 10^{-1}$ & $7.79 \times 10^{-2}$ & $2.00 \times 10^{-2}$ & $1.81 \times 10^{-3}$ & $6.38 \times 10^{-4}$ & $3.99 \times 10^{-4}$\\ 
&$E_{shape}$ & $1.37 \times 10^{-1}$ & $4.89 \times 10^{-2}$ & $1.88 \times 10^{-2}$ & $7.07 \times 10^{-3}$ & $3.60 \times 10^{-3}$ & $1.77 \times 10^{-3}$\\
&$E_{sym}$ &$4.13 \times 10^{-2}$ & $2.11 \times 10^{-2}$ & $8.09 \times 10^{-3}$ & $3.27 \times 10^{-3}$ & $1.56 \times 10^{-3}$ & $7.26 \times 10^{-4}$\\
\hline
Unsplit-Euler &$E_{area}$ & $2.46 \times 10^{-1}$ & $8.31 \times 10^{-2}$ & $1.96 \times 10^{-2}$ & $1.38 \times 10^{-3}$ & $9.90 \times 10^{-5}$& $2.06 \times 10^{-4}$\\
&$E_{shape}$ & $1.38 \times 10^{-1}$ & $6.37 \times 10^{-2}$ & $3.21 \times 10^{-2}$ & $1.62 \times 10^{-2}$ & $8.12 \times 10^{-3}$ & $4.07 \times 10^{-3}$\\
&$E_{sym}$ & $6.62 \times 10^{-2}$ & $3.42 \times 10^{-2}$ & $1.56 \times 10^{-2}$ & $7.28 \times 10^{-3}$ & $3.59 \times 10^{-3}$ & $1.79 \times 10^{-3}$\\
\hline 
Unsplit-PC &$E_{area}$ & $4.61 \times 10^{-1}$ & $1.33 \times 10^{-1}$ & $1.87 \times 10^{-2}$ & $2.64 \times 10^{-3}$ & $7.31 \times 10^{-4}$ & $2.40 \times 10^{-4}$\\ 
&$E_{shape}$ & $1.19 \times 10^{-1}$ & $4.16 \times 10^{-2}$ & $1.31 \times 10^{-2}$ & $2.45 \times 10^{-3}$ & $1.05 \times 10^{-3}$ & $5.58 \times 10^{-4}$\\
&$E_{sym}$ & $4.40 \times 10^{-2}$ & $1.02 \times 10^{-2}$ & $1.65 \times 10^{-3}$ & $2.52 \times 10^{-4}$ & $6.94 \times 10^{-5}$ & $2.39 \times 10^{-5}$\\
\hline
Unsplit-RK4 &$E_{area}$ & $4.15 \times 10^{-1}$ & $1.15 \times 10^{-1}$ & $1.72 \times 10^{-2}$ & $2.74 \times 10^{-3}$ & $7.38 \times 10^{-4}$ & $2.39 \times 10^{-4}$\\ 
&$E_{shape}$ & $1.21 \times 10^{-1}$ & $5.44 \times 10^{-2}$ & $1.34 \times 10^{-2}$& $2.53 \times 10^{-3}$ & $1.06 \times 10^{-3}$ & $5.56 \times 10^{-4}$\\
&$E_{sym}$ & $3.46 \times 10^{-2}$ & $1.14 \times 10^{-2}$ & $1.61 \times 10^{-3}$& $2.55 \times 10^{-4}$ & $7.00 \times 10^{-5}$ & $2.39 \times 10^{-5}$\\
\hline 
\end{tabular}
\label{Tab_vortex_error_t8}
\normalsize
\end{table}
% ----------

In Figs.~\ref{Fig_vortex_intf_t2_1} and \ref{Fig_vortex_intf_t2_2} at half time,
the interface lines are highly stretched but still retain their integrity.
This is not true for the $T=8$ simulation, where the interface is not adequately
resolved and artificial fragmentation occurs. For this reason, we compare
the new results with those obtained with the VOF method of the Basilisk platform.
The shape and symmetric difference errors, $E_{shape}$ and $E_{sym}$, respectively,
of Fig.~\ref{Fig_vortex_error_t2} show second-order convergence with grid refinement
for the unsplit-PC and unsplit-RK4 methods and the VOF method. Moreover, the errors
of the unsplit methods are systematically somewhat smaller.

For the test with period $T = 8$, the interface lines at maximum deformation and
the end of the simulation are shown in Figs.~\ref{Fig_vortex_intf_t8_1} and
\ref{Fig_vortex_intf_t8_2} for different mesh resolutions. The interface line at
half time is stretched into a long, thin ligament, completing almost two spiral
turns. When the mesh resolution is too coarse, that is, when $N_x$ is equal to $32$ 
and $64$, fragmentation of the interface at half time and strong deformation
at the end of the simulation are observed for all methods. 
As the mesh resolution is increased, convergence to the reference solution is
always found, much faster for the unsplit-PC and unsplit-RK4 methods than
for the split and unsplit-Euler methods. Moreover, the deviation from the
reference solution of the unsplit-Euler method for this test is much more
pronounced than that of the other methods.

% ----------
\begin{figure}
\begin{center}
\begin{tabular}{ccc}
\includegraphics[width=0.33\textwidth]{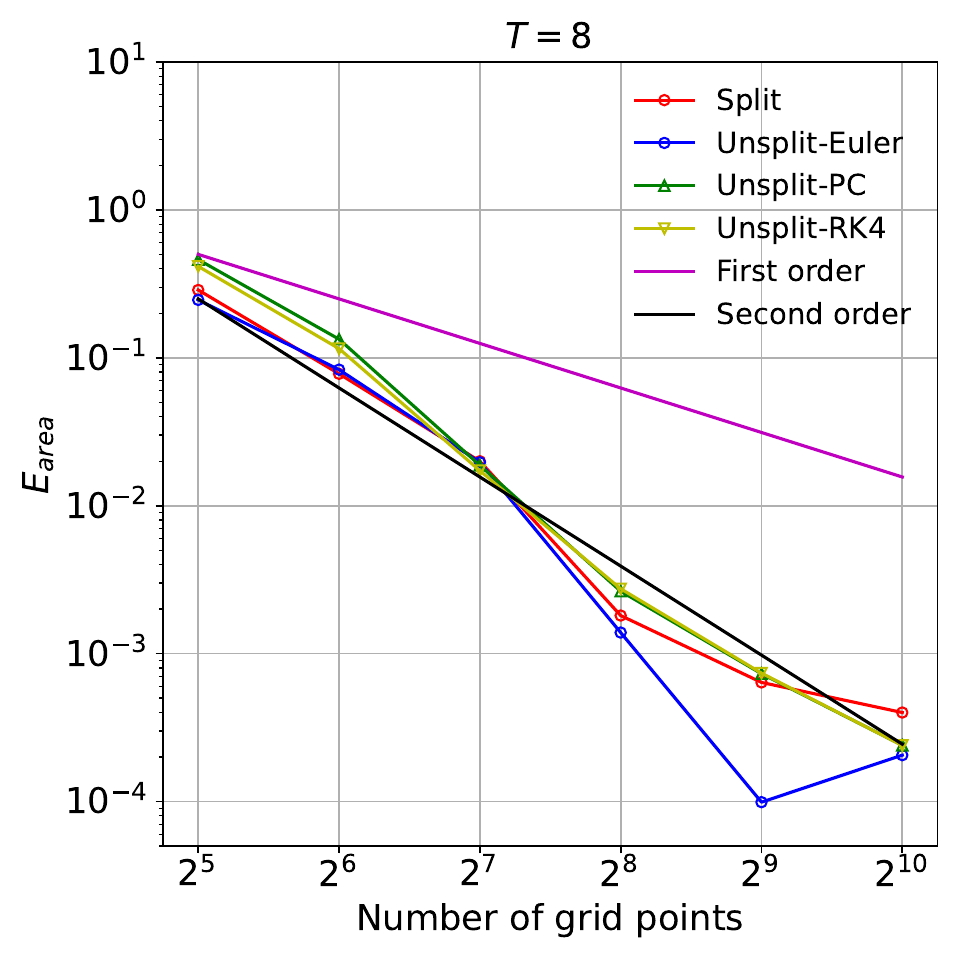} &
\includegraphics[width=0.33\textwidth]{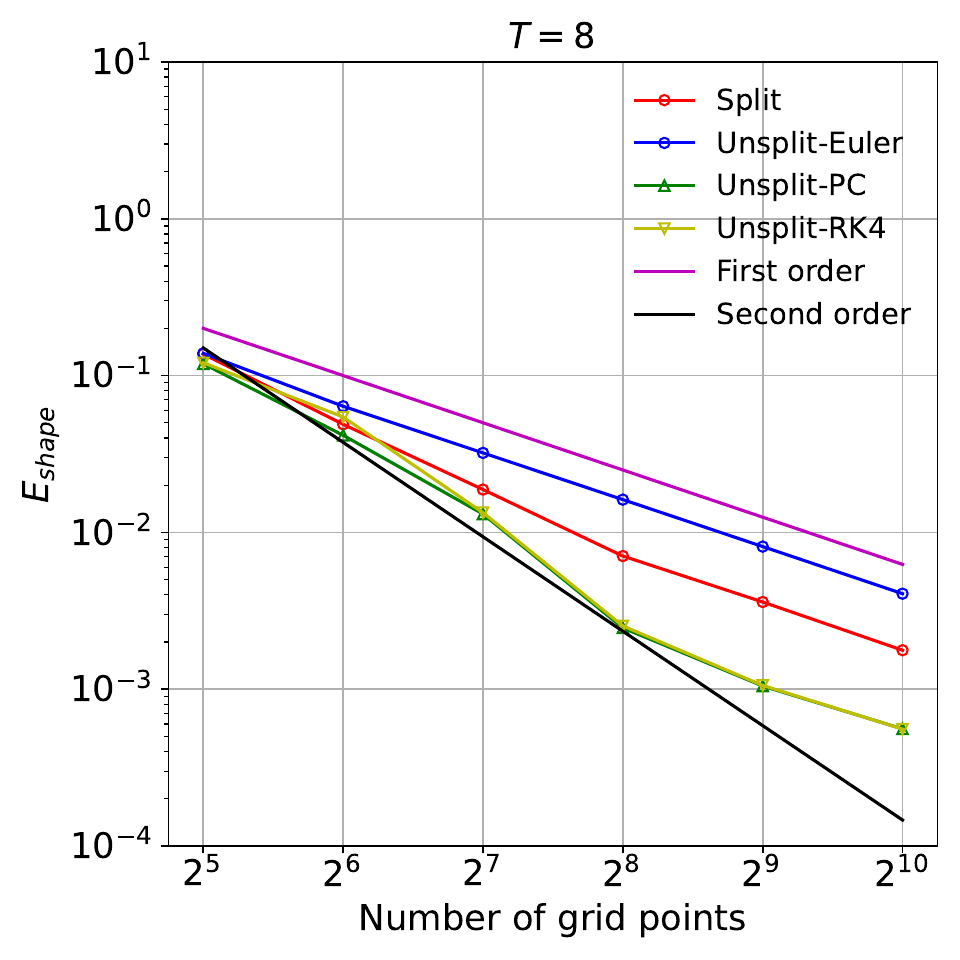}&
\includegraphics[width=0.33\textwidth]{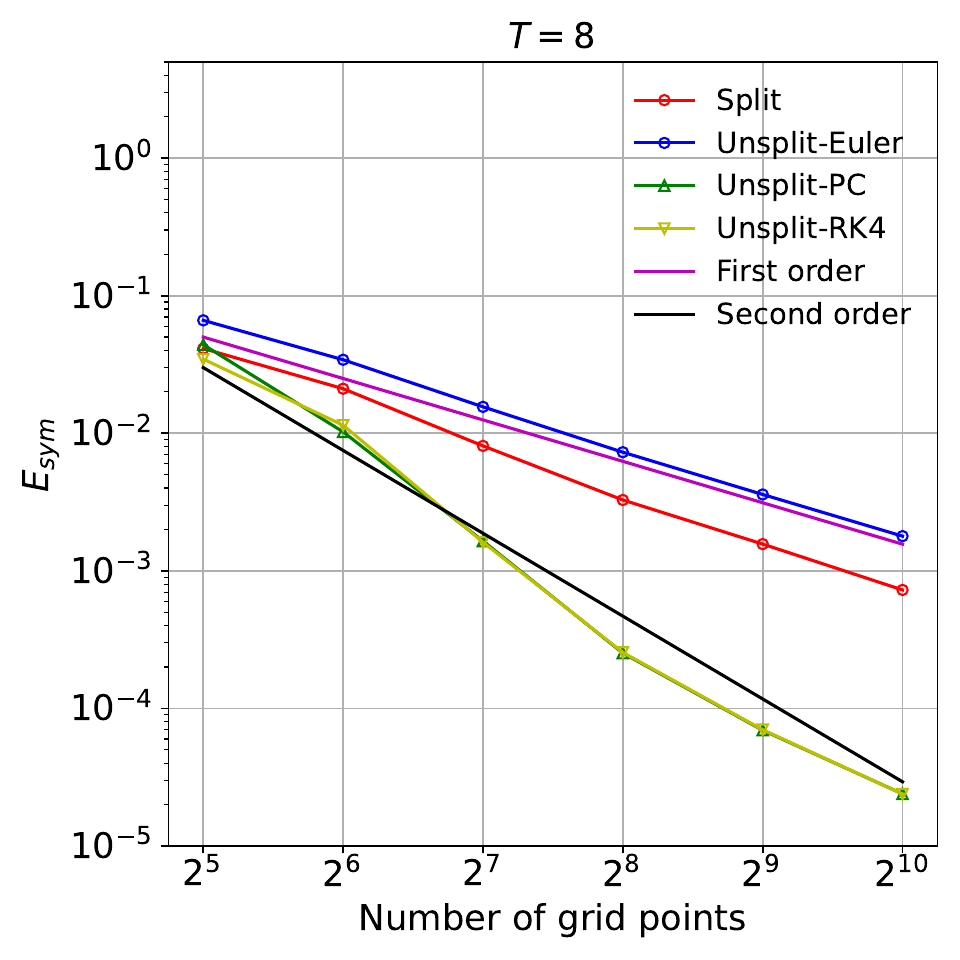}\\
(a) & (b) & (c)
\end{tabular}
\end{center}
\caption{Errors in the single vortex test with period $T = 8$ for different methods as a function of grid resolution: (a) area error $E_{area}$; (b) shape error $E_{shape}$; (c) symmetric difference error $E_{sym}$}
\label{Fig_vortex_error_t8}
\end{figure}
% ----------

The area error $E_{area}$, the shape error $E_{shape}$, and the symmetric difference
error $E_{sym}$ are listed in Table~\ref{Tab_vortex_error_t8} and are shown in
Fig.~\ref{Fig_vortex_error_t8}.  A very high resolution is required to stabilize 
the order of convergence of the three errors, $N_x = 1024$.

For all methods considered, second-order convergence is observed for the area error. 
The fluctuation of the area error for the unsplit-Euler method is probably caused by
a sign change of the error. At low resolutions, there is a mass loss, $A(T) < A(0)$,
at high resolutions, a mass gain, $A(T) > A(0)$. 
 
For the split and unsplit-Euler methods, first-order convergence is observed for the
shape and symmetric difference errors. These two errors are systematically smaller
for the split method.

For the unsplit-PC and unsplit-RK4 methods, second-order convergence is observed
for the symmetric difference error and first-order convergence for the shape error.
The results obtained with these two unsplit schemes for the single vortex test, with
two different values of the period T, are very close, even if the first
scheme is second-order accurate, while the second one is a fourth-order method.
This is a clear indication that the dominant error is the reconstruction step,
based on a circular fit.

\section{Conclusions}

We present an unsplit scheme for the interface advection in our novel Front-Tracking method,
the Edge-Based Interface Tracking (EBIT) method. The algorithms for reconstructing the
interface and for updating the color vertex field adapted to the unsplit scheme are
presented in detail. 
The unsplit scheme facilitates the implementation of high-order time integration methods.
Thus, two high-order methods, the second-order Predictor-Corrector (PC) method, and the
fourth-order Runge-Kutta (RK4) method, are coupled with the unsplit scheme and
investigated.

The unsplit EBIT method has been implemented in the free Basilisk platform. Several
kinematic test cases are considered to validate the unsplit scheme and to compare it
with the split scheme used in our original version of the EBIT method.

For the two methods based on the first-order explicit Euler method for time integration,
the split method is generally more accurate than the unsplit-Euler method, in terms
of area, shape, and symmetric errors, since a correction step is implicitly included
in the split method.

The accuracy of the unsplit-PC and unsplit-RK4 methods is related to the velocity
field in the computational domain. 
These two methods provide more accurate results than the split and unsplit-Euler methods
for a smooth interface. 
For a velocity field with a linear spatial distribution, an order of convergence greater
than two is observed for both of them. However, for nonlinear velocity fields, 
such as the single vortex test, second-order convergence is observed for the symmetric
difference error. The errors are dominated by the reconstruction step, and the results
obtained with these two unsplit time integration methods are very close.

In addition to the potential for employing more accurate time integration methods, the
unsplit scheme also facilitates a more straightforward coupling between the EBIT method 
and boundary layer models for multiscale problems. The EBIT method is merged with
the quad/octree structure of the Basilisk platform, with proven scalability on 
high-performance computers. Thus, there is hope that its coupling with boundary models 
and its automatic parallelization capability can bring progress in multiscale 
simulations.

\section{CRediT authorship contribution statement}

\textbf{J. Pan}: Writing - review \& editing, Writing - original draft, Validation, 
Software, Methodology, Formal analysis, Data curation, Conceptualization.

\textbf{T. Long}: Software, Methodology, Formal analysis, Conceptualization.

\textbf{R. Scardovelli}: Writing - review \& editing, Writing - original draft, 
Software, Methodology, Formal analysis,  Conceptualization.

\textbf{S. Zaleski}: Writing - review \& editing, Writing - original draft, 
Supervision, Resources, Project administration, Funding acquisition, 
Methodology, Formal analysis, Conceptualization.

\section{Declaration of competing interest}
The authors declare that they have no known competing financial interests or personal relationships that could have appeared to influence the work reported in this paper.

\section{Acknowledgements}
St\'{e}phane Zaleski recalls meeting Sergei Semushin in March 1995 and learning about his method. He thanks him for the extensive and fruitful discussions about the method.
This project has received funding from the European Research Council (ERC) under the European Union's Horizon 2020 research and innovation programme (grant agreement number 883849).

%\appendix

%\section{Section in Appendix}
%\label{appendix-sec1}

%Sample text. Sample text. Sample text. Sample text. Sample text. Sample text. 
%Sample text. Sample text. Sample text. Sample text. Sample text. Sample text. 
%Sample text. 

%% References


%\bibliographystyle{elsarticle-num}
% \bibliographystyle{elsarticle-harv}
% \bibliographystyle{elsarticle-num-names}
% \bibliographystyle{model1a-num-names}
% \bibliographystyle{model1b-num-names}
% \bibliographystyle{model1c-num-names}
\bibliographystyle{model1-num-names}
\bibliography{multiphase-jieyun,multiphase-stephane}
\end{document}